\documentclass[aps,showpacs,showkeys,amsmath,amssymb,twocolumn,prc,%
floatfix,superscriptaddress]{revtex4}

\usepackage{graphicx}
\usepackage{dcolumn}% Align table columns on decimal point
\usepackage{docs}%
\usepackage{bm}% bold math
\begin{document}
%%%

\title{A Family of Equations of State Based on Lattice QCD:\\
Impact on Flow in Ultrarelativistic Heavy-Ion Collisions}
%%%
\author{M.~Bluhm}
\affiliation{%Institut f\"ur Kern- und Hadronenphysik, 
  Forschungszentrum Dresden-Rossendorf, PF 510119, 01314 Dresden, Germany}
\author{B.~K\"ampfer}
\affiliation{%Institut f\"ur Kern- und Hadronenphysik, 
  Forschungszentrum Dresden-Rossendorf, PF 510119, 01314 Dresden, Germany}
\affiliation{Institut f\"ur Theoretische Physik, TU Dresden, 
01062 Dresden, Germany}
\author{R. Schulze}
\affiliation{Institut f\"ur Theoretische Physik, TU Dresden, 
01062 Dresden, Germany}
\author{D. Seipt}
\affiliation{Institut f\"ur Theoretische Physik, TU Dresden, 
01062 Dresden, Germany}
\author{U. Heinz}
\affiliation{Department of Physics, The Ohio State University, 
Columbus, OH 43210, USA}

\date{\today}

%Use showkeys class option if keyword display desired
%PACS, the Physics and Astronomy Classification Scheme.

%%%%%%%%%%%%%%%%%%%%%%%%%%%%%%%%%%%%%%%%%%%%%%%%%%%%%%%%%%%%%%%%%%%%%%%%%%%%%%
\begin{abstract}
We construct a family of equations of state within a quasiparticle model 
by relating pressure, energy density, baryon density and susceptibilities 
adjusted to first-principles lattice QCD calculations. The relation 
between pressure and energy density from lattice QCD is surprisingly 
insensitive to details of the simulations. Effects from different
lattice actions, quark masses and lattice spacings used in the simulations
show up mostly in the quark-hadron phase transition region which we bridge 
over by a set of interpolations to a hadron resonance gas equation of state.
Within our optimized quasiparticle model we then examine the equation of 
state along isentropic expansion trajectories at small net baryon densities,
as relevant for experiments and hydrodynamic simulations at RHIC and LHC
energies. We illustrate its impact on azimuthal flow anisotropies and 
transverse momentum spectra of various hadron species.
\end{abstract}
%%%%%%%%%%%%%%%%%%%%%%%%%%%%%%%%%%%%%%%%%%%%%%%%%%%%%%%%%%%%%%%%%%%%%%%%%%%%%%

\maketitle

\section{Introduction \label{sec:Intro}}

In the last few years, much evidence has been accumulated for the 
applicability of hydrodynamics in describing the expansion stage of 
strongly interacting matter created in relativistic heavy-ion collisions 
\cite{KSH,TLS,Shu,Huovinen,Huov05,Heinz_SQM04,WhitePapers}. Hydrodynamics 
describes the collective flow of bulk matter from an initial state 
just after reaching thermalization up to the kinetic freeze-out stage. 
The heart of hydrodynamics is the equation of state (EoS) which relates 
thermodynamically the pressure $p$ of the medium to its energy density $e$
and net baryon density $n_B$ (or, equivalently, to its temperature $T$ 
and baryon chemical potential $\mu_B$). Specifically, the parameter 
controlling the acceleration of the fluid, i.e. the build-up of collective 
flow, by pressure gradients in the system is the speed of sound, given 
by $c_s^2=\frac{\partial p}{\partial e}$. 

While most existing hydrodynamic simulations have used a realistic 
hadron resonance gas EoS below the deconfinement transition (either with full 
\cite{KSH,Huovinen,Huov05} or partial 
\cite{TLS,Hirano01,Teaney_chem,Rapp_chem,KR03} chemical 
equilibrium among the hadron species), they have usually relied on
simple analytical models for the EoS of the quark-gluon plasma (QGP) 
above the transition, based on the assumption of weak coupling among the 
deconfined quarks and gluons. This assumption is, however, inconsistent
with the phenomenological success of hydrodynamics which requires rapid
thermalization of the QGP \cite{SQCD} and therefore strong interactions 
among its constituents \cite{PANIC05,Gyu_Kemer,GyMcL05,Shu04}. Indeed,
lattice QCD calculations of the QGP pressure and energy density
show that they deviate from the Stefan-Boltzmann limit for an ideal 
gas of non-interacting quarks and gluons even at temperatures $T>3\,T_c$
(with $T_c$ as pseudo-critical temperature), 
by about 15-20\% \cite{Kar1,Karsch_review,Karsch_QGP3}. Miraculously,
however, the deviations are of similar magnitude in both $p$ and $e$
such that, for $T\gtrsim 2\,T_c$, the squared speed of sound 
$c_s^2=\frac{\partial p}{\partial e}\approx\frac{1}{3}$ 
\cite{Karsch_QGP3}, just as expected for a non-interacting gas of 
massless partons. In spite of the evidence for strong interactions among
the quarks and gluons in the QGP seen in both $p(T)$ and $e(T)$, the
stiffness and accelerating power of the lattice QCD equation of state
is thus indististinguishable from that of an ideal parton gas (at least
for temperatures $T\gtrsim 2\,T_c$), such as the one used above $T_c$ in 
most hydrodynamical simulations.

On the other hand, at $T < 2\,T_c$ the speed of sound extracted
from lattice QCD drops below the ideal gas value $c_s=1/\sqrt{3}$,
reaching a value that is about a factor of 3 smaller near $T_c$
\cite{Karsch_QGP3}. This leads to a significant softening of the
QGP equation of state relative to that of an ideal massless gas
exactly in the temperature region $T_c{\,<\,}T{\,<\,}2\,T_c$ explored
during the early stages of Au+Au collisions at RHIC 
\cite{KSH,TLS,Huovinen,Huov05,Hirano01}. To explore the
sensitivity of the flow pattern seen in the RHIC data to such details
of the EoS near the quark-hadron phase transition, the hydrodynamic
evolution codes must be supplied with an EoS that faithfully reproduces
the lattice QCD results above $T_c$. To construct such an EoS, and to 
test its influence on the collective flow generated in RHIC and LHC 
collisions, are the main goals of this paper.

Our approach is based on the quasiparticle model \cite{Peshier,Levai,%
Schneider,Letessier,Rebhan,Thaler,Rischke,Ivanov,Bannur,Peshier05} which 
expresses the thermodynamic quantities as standard phase space integrals over 
thermal distribution functions for quasiparticles with medium dependent 
properties. In the present paper we follow the philosophy 
\cite{Peshier,Levai,Schneider,Letessier,Rebhan,Thaler,Rischke,Ivanov,Bannur} 
that the interaction effects in the QGP can be absorbed into the 
quasiparticle masses and a vacuum energy all of which depend on the 
temperature and baryon chemical potential. This is known to produce 
good fits to the lattice QCD data both at vanishing 
\cite{Peshier,Levai,Schneider,Rebhan} and non-vanishing 
\cite{Letessier,Thaler,Ivanov} baryon chemical potential. However,
since this approach uses on-shell spectral functions for the 
quasiparticles, it implicitly assumes zero residual interactions 
(i.e. infinite mean free paths) for them, which is inconsistent 
with the low viscosity and almost ideal fluid dynamical behaviour of 
the QGP observed at RHIC. Peshier and Cassing \cite{Peshier05} have
shown that it is possible to generalize the quasiparticle description
to include a finite (even large) collisional width in the spectral 
functions, without significantly affecting the quality of the model fit to 
the lattice QCD data for the EoS at $\mu_B=0$. Since hydrodynamics only 
cares about the EoS, but not about its microscopic interpretation, we
here opt for the simpler, but equally successful approach using on-shell
quasiparticles to fit the lattice QCD EoS.

The quasiparticle EoS for the QGP above $T_c$ does not automatically
match smoothly with the hadron resonance gas EoS below $T_c$. Although
the gap between the two branches of the EoS is much smaller here than for 
the previously used models which assume non-interacting quarks and gluons 
above $T_c$ \cite{KSH,Huovinen,Shu,Huov05,TLS,Hirano01,Teaney_chem,%
Rapp_chem,KR03}, a certain degree of ambiguity remains in the 
interpolation process. We explore a set of different interpolation 
prescriptions, yielding a family of equations of state which exhibit 
slight differences in the phase transition region, and study their 
dynamical consequences.

Our paper is organized as follows: In Sec.~\ref{sec:QPM} we show that 
our quasiparticle model provides an efficient and accurate parametrization 
of lattice QCD results for $N_f{\,=\,}2$ flavors both at $\mu_B=0$ and 
$\mu_B\ne0$. We also extract the isentropic expansion trajectories 
followed by fully equilibrated systems. In that Section, the quasiparticle 
parametrization is continued below $T_c$, down to temperatures of about 
$0.75\,T_c$ where the lattice QCD data end. In Sec.~\ref{sec:FamilyEoS} 
we proceed to the physically relevant case of $N_f = 2 + 1$ flavors and 
furthermore match the quasiparticle EoS above $T_c$ to a hadron resonance 
gas EoS below $T_c$. Variations in the matching procedure lead to a family 
of equations of state with slightly different properties near $T_c$. The 
transition to a realistic hadron resonance gas picture below $T_c$ means 
that these EoS can now be used down to much lower temperatures to make 
explicit contact with the experimentally observed final state hadrons 
after decoupling from the expanding fluid. In Sec.~\ref{sec:ellipticflow} 
we use this family of EoS for hydrodynamic calculations of the 
differential elliptic flow $v_2(p_T)$ for several hadronic species 
in Au+Au collisions at the top RHIC energy and compare with experimental 
data. We find some sensitivity to the details of the interpolation 
scheme near $T_c$, as long as an EoS is used that agrees with the 
lattice QCD data for energy densities $e>4$\,GeV/fm$^3$. We conclude 
that Section with a few predictions for Pb+Pb collisions at the LHC.
A short summary is presented in Sec.~\ref{sec:conclusions}. 

%%%%%%%%%%%%%%%%%%%%%%%%%%%%%%%%%%%%%%%%%%%%%%%%%%%%%%%%%%%%%%%%%%%%%
\section{Quasiparticle description of the EOS from lattice QCD for 
$\bm{N_f=2}$ 
\label{sec:QPM}}
%%%%%%%%%%%%%%%%%%%%%%%%%%%%%%%%%%%%%%%%%%%%%%%%%%%%%%%%%%%%%%%%%%%%%
\subsection{The quasiparticle model}
%%%%%%%%%%%%%%%%%%%%%%%%%%%%%%%%%%%%%%%%%%%%%%%%%%%%%%%%%%%%%%%%%%%%%

Over the years, several versions of quasiparticle models have been 
developed to describe lattice QCD data for the QCD equation of state
\cite{Peshier,Levai,Schneider,Letessier,Rebhan,Thaler,Rischke,Ivanov,%
Peshier05}. They differ in the choice and number of parameters and
in the details of the underlying microscopic picture but generally 
yield fits to the lattice QCD data which are of similar quality. In
this subsection we quickly review the essentials of the model 
described in \cite{Peshier} which will be used here.

In our quasiparticle approach the thermodynamic pressure is written as 
a sum of contributions associated with medium modified light quarks $q$,
strange quarks $s$, and gluons $g$ \cite{Peshier}: 
\begin{equation}
  \label{e:pres0}
  p (T,\{\mu_a\}) = \sum_{a = q,s,g} \!\!\! p_a - B(T, \{\mu_a\}) \,,
\end{equation}
with partial pressures 
\begin{equation}
  \label{e:pres1}
  p_a = \frac{d_a}{6 \pi^2} \int_0^\infty dk 
        \frac{k^4}{\omega_a}\left( f_a^+ + f_a^- \right)\,.
\end{equation}
Here $f_a^\pm = (\exp( [\omega_a \mp \mu_a]/ T) +S_a)^{-1}$ are thermal
equilibrium distributions for particles and antiparticles, with 
$S_{q,s}=1$ for fermions and $S_g=-1$ for bosons. $d_a$ represents the
spin-color degeneracy factors, with $d_q=2N_qN_c=12$ for the $N_q=2$ 
light quasi-quarks, $d_s=2N_c=6$ for the strange quasi-quarks, and 
$d_g=N_c^2-1=8$ for the right-handed
transversal quasi-gluons (with the left-handed ones counted as their 
antiparticles). Since the pressure integral in Eq.~(\ref{e:pres1})
is dominated by thermal momenta of order $k\sim T$, weak coupling 
perturbation theory suggests \cite{leBellac,Kapusta} that the dominant 
propagating modes are transversal plasmons with gluon quantum numbers 
($g$) and quark-like excitations, whereas longitudinal plasmons are 
exponentially suppressed. Our model assumes that this remains true near
$T_c$ where perturbation theory is not expected to be valid.

We are interested in the application of this EoS to heavy ion collisions
where strangeness is conserved at its initial zero value, due to the very
short available time. This strangeness neutrality constraint allows to
set $\mu_s=0$. The isospin chemical potential $\mu_I=(\mu_u-\mu_d)/2$ is 
fixed by the net electric charge density of the medium; we assume zero 
net charge of the fireball matter created near midrapidity at RHIC as 
well as equal masses for the $u$ and $d$ quasi-quarks such that 
$\mu_I=0$ and we have only one independent chemical potential 
$\mu_u=\mu_d\equiv\mu_q=\mu_B/3$ where $\mu_B$ is the baryon number 
chemical potential.

The quasiparticles are assumed to propagate on-shell, i.e, with real
energies $\omega_a$ given by dispersion relations of the type 
$\omega_a = \sqrt{k^2 + m_a^2(T,\mu_q)}$, known to hold for weakly 
interacting quarks and gluons with thermal momenta $k \sim T$. Again the 
model assumes that this structure holds true also near $T_c$ where 
perturbation theory presumably breaks down. In order to directly compare 
our Quasiparticle Model (QPM) with lattice QCD results, we include nonzero 
bare quark masses $m_{a0}$ and adjust them to the values used in the 
lattice simulations through $m_a^2 = m_{a0}^2 + \Pi_a$ \cite{Pisarski}
where $\Pi_a$ denotes the self energy. For gluonic modes we use
$m_{g0} = 0$. For $\Pi_a$ we employ an ansatz inspired by the 
asymptotic form of the gauge independent hard thermal/dense loop 
(HTL/HDL) self-energies which depend on $T$, $\mu_q$, $m_{a0}$, 
and the running coupling $g^2$ as follows \cite{leBellac,Pisarski}: 
\begin{eqnarray}
  \label{e:pig}
  \Pi_g & = & \left(\left[3+\frac{N_f}{2}\right]T^2 + 
  \frac{3}{2\pi^2}\sum_f \mu_f^2\right)\frac{g^2}{6} , \\
  \label{e:piq}
  \Pi_q & = & 2 m_{q0} \sqrt{\frac{g^2}{6}
              \left(T^2+\frac{\mu_q^2}{\pi^2}\right)}+
  \frac{g^2}{3}\left(T^2+\frac{\mu_q^2}{\pi^2}\right) , \\
  \label{e:pis}
  \Pi_s & = & 2 m_{s0} \sqrt{\frac{g^2}{6} T^2}+
  \frac{g^2}{3} T^2 \, . 
\end{eqnarray}
The $m_a$ in the dispersion relations thus denote effective quasiparticle 
masses due to the dynamically generated self-energies $\Pi_a$. The mean 
field interaction term $B(T,\mu_q)$ in Eq.~(\ref{e:pres0}) is determined 
by thermodynamic self-consistency and stationarity of the thermodynamic 
potential under functional variation of the self-energies, 
$\delta p / \delta \Pi_a = 0$ \cite{Gorenstein}. As a consequence, 
$B(T,\mu_q)$ is evaluated in terms of an appropriate line integral in 
the $T$-$\mu_q$ plane, with integration constant $B(T_c)$ adjusted 
to the lattice results \cite{Peshier}. 

All other thermodynamic quantities follow straightforwardly from the 
stationarity condition and standard thermodynamic relations. For example, 
the entropy density reads 
\begin{equation}
  \label{e:entr0}
  s = \sum_{a = q,s,g} s_a,
\end{equation}  
\vspace*{-3mm}
\begin{equation*}
%  \label{e:entr1}
  s_a = \frac{d_a}{2\pi^2} \int_0^\infty \!\!\! k^2 dk
  \Bigl[ \frac{\left( \frac{4}{3}k^2{+}m_a^2 \right)}{\omega_a T}
    (f_a^+{+}f_a^-) - \frac{\mu_a}{T} (f_a^+{-}f_a^-) \Bigr]
\end{equation*}
while the net quark number density $n_q=3\,n_B$ is given through
\begin{equation}
  \label{e:dens1}
  n_q = \frac{d_q}{2\pi^2} \int_0^\infty \!\!\! k^2 dk (f_q^+ - f_q^-) .
\end{equation}

Although the form of our ansatz for the quasiparticle masses (i.e. the
specific interplay between the parameters $m_{a0}, T,$ and $\mu_q$) is 
inspired by perturbation theory, our model becomes non-perturbative
by replacing the perturbative expression for the running coupling 
$g^2$ in Eqs.~(\ref{e:pig}-\ref{e:pis}) by an effective coupling $G^2$ 
whose dependence on $T$ and $\mu_q$ is parametrized and fitted to the 
non-perturbative $(T,\mu_q)$-dependence of the thermodynamic functions 
from lattice QCD. The $(T,\mu_q)$-dependence of $G^2$ is constrained by 
Maxwell's relation for $p$ which takes the form of a quasi-linear 
partial differential equation 
\begin{equation} 
  \label{e:flow}
  a_{\mu_q} \frac{\partial G^2}{\partial \mu_q} +
  a_T \frac{\partial G^2}{\partial T} = b;
\end{equation}
here $a_{\mu_q}$, $a_T$ and $b$ depend on $T$, $\mu_q$ and $G^2$ 
(see Refs.~\cite{Peshier,Bluhm} for details). This flow equation is
solved by the method of characteristics, starting from initial conditions
on a Cauchy surface in the $T$-$\mu_q$ plane. One possibility is to 
parameterize $G^2$ at $\mu_q=0$ such that lattice QCD results for 
vanishing quark chemical potential are reproduced, and to use the flow
equation for extrapolation to non-zero $\mu_q$. As a convenient 
parametrization of $G^2(T,\mu_q{=}0)$ we find \cite{Bluhm04}
\begin{equation}
  \label{e:G2param}
  G^2(T,\mu_q{=}0) = \left\{
    \begin{array}{l}
      \!\! G^2_{\rm 2-loop} (T), \quad T{\,\ge\,}T_c,
      \\[3mm]
      \!\! G^2_{\rm 2-loop}(T_c) + b \left(1{-}\frac{T}{T_c}\right), \ 
      T{\,<\,}T_c .
    \end{array}
  \right.\!\!\!
\end{equation}
Here, in order to recover perturbation theory in the high temperature limit,
$G^2_{\rm 2-loop}$ is taken to have the same form as the perturbative
running coupling at 2-loop order:
\begin{equation}
  \label{e:G2pert}
  G^2_{\rm 2-loop}(T) = \frac{16 \pi^2}{\beta_0 \log \xi^2}
  \left[1 - \frac{2 \beta_1}{\beta_0^2} \frac{\log (\log \xi^2)}{\log \xi^2} 
  \right],
\end{equation}
with $\beta_0 = \frac{1}{3}(11 N_c - 2 N_f)$ and $\beta_1 = \frac{1}{6}
(34 N_c^2 -13 N_f N_c + 3 N_f /N_c)$. The scale $\xi$ is parametrized
phenomenologically as $\xi = \lambda (T - T_s)/T_c$, with a scale 
parameter $\lambda$ and a temperature shift $T_s$ which regulates the 
infrared divergence of the running coupling by shifting it somewhat
below the critical temperature $T_c$. Below the phase transition, 
we postulate a continuous linear behavior of the effective coupling. 
The parametrization (\ref{e:G2param},\ref{e:G2pert}) turns out to be 
flexible enough to describe the lattice QCD results accurately down to 
about $T\approx0.75\,T_c$. In contrast, using a pure 1-loop or 2-loop 
perturbative coupling together with a more complete description of the 
plasmon term and Landau damping restricts the quasiparticle approach to 
$T > 2\,T_c$~\cite{Blaizot}. [Similar quality fits can be achieved in that
approach, without giving up its more accurate form of the HTL/HDL self
energies, by adopting a similar non-perturbative modification of 
the running coupling as adopted here \cite{Rebhan}.] 

The model described in this subsection was successfully applied to 
QCD lattice data in the pure gauge sector in Ref.~\cite{Peshier}, and 
to first lattice QCD calculations at $\mu_q\ne 0$ in Ref.~\cite{Szabo}.
In the following subsection we test it on recent lattice QCD data for 
$N_f=2$ dynamical quark flavors at zero and non-zero $\mu_q$, and in 
the next section we consider the realistic case of $N_f=2+1$ flavors with 
the aim of providing an EoS suitable for hydrodynamic simulations of 
heavy-ion collisions. 

%%%%%%%%%%%%%%%%%%%%%%%%%%%%%%%%%%%%%%%%%%%%%%%%%%%%%%%%%%%%%%%%%%%%%%%%%%%%
\subsection{Thermodynamics of \boldmath{$N_f=2$} quark flavors 
\label{sec:QPMcompNf2}}
%%%%%%%%%%%%%%%%%%%%%%%%%%%%%%%%%%%%%%%%%%%%%%%%%%%%%%%%%%%%%%%%%%%%%%%%%%%%

\begin{figure}[ht]
\includegraphics[bb=80 30 570 700,scale=0.35,angle=-90.,clip=]{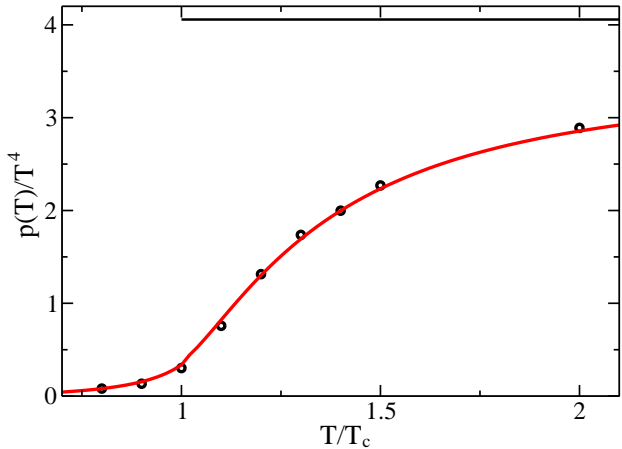}
\caption{(Color online) Comparison of the QPM with lattice QCD results (symbols) for the 
         scaled pressure $p/T^4$ as a function of $T/T_c$ for $N_f=2$ 
         and $\mu_q=0$. The raw lattice QCD data from \cite{Kar1} have 
         been continuum extrapolated as described in the text. The QPM 
         parameters are $\lambda = 4.4$, $T_s=0.67T_c$, $b=344.4$ and 
         $B(T_c)=0.31 T_c^4$, with $T_c=175$ MeV as suggested in 
         \cite{Ejiri}. The horizontal line indicates the Stefan-Boltzmann 
         value $p_\mathrm{SB}/T^4=\bar{c}_0=(32+21N_f)\pi^2/180$ for $N_f=2$. 
    \label{fig:2}}
\end{figure}

We begin with the case of $N_f=2$ dynamical quark flavors at zero quark 
chemical potential and confront the QPM with lattice QCD results 
obtained by the Bielefeld-Swansea collaboration \cite{Kar1}. These 
simulations were performed with temperature dependent bare quark 
masses $m_{a0}(T)=x_a T$ where $x_g=0$ and $x_q=0.4$ \cite{Kar1}. For
$N_f=2$ light quark flavors we can set $d_s=0$ in the QPM expressions.
Fig.~\ref{fig:2} shows the lattice QCD data for the scaled pressure 
$p(T)/T^4$ together with the QPM fit; the fit parameters given in the 
caption were obtained by the procedure described in Ref.~\cite{Bormio}.
The raw lattice data were extrapolated to the continuum by multiplying
the pressure in the region $T\ge T_c$ by a constant factor $d=1.1$, following
an estimate given in \cite{Kar1,Kar3} who advocate a range of 10-20\% due 
to finite size and cutoff effects. (Note that this estimated correction 
factor does not necessarily have to be independent of $T$, as assumed here.)

%
%%%%%%%%%%%%%%%%%%%%%%%%%%% Fig. 1 %%%%%%%%%%%%%%%%%%%%%%%%%%%%%%%%%%%%%%%%%%%
%%%%%%%%%%%%%%%%%%%%%%%%%%%%%%%%%%%%%%%%%%%%%%%%%%%%%%%%%%%%%%%%%%%%%%%%%%%%%%
%
Having demonstrated the ability of the QPM to successfully reproduce
lattice EoS data along the $\mu_q=0$ axis, we can now exploit recent 
progress in lattice QCD with small non-vanishing chemical potential to 
test its ability to correctly predict the thermodynamic functions at 
non-zero $\mu_q$. In Ref.~\cite{All05} finite-$\mu_q$ effects were 
evaluated by expanding the pressure into a Taylor series in powers of 
$(\mu_q/T)$ around $\mu_q=0$,
\begin{equation}
  \label{e:presdecomp}
  p(T, \mu_q) = T^4 \sum_{n=0,2,4,\dots}^\infty c_n(T) 
  \left( \frac{\mu_q}{T} \right)^n ,
\end{equation}
where $c_0(T)=p(T,\mu_q{=}0)/T^4$ is the scaled pressure at vanishing 
quark chemical potential. The coefficients $c_2(T),\,c_4(T),\,c_6(T)$ 
were extracted from the lattice by numerically evaluating appropriate
$\mu_q$-derivatives of the logarithm of the partition function 
$\ln Z = pV/T$ \cite{All05}, viz.
\begin{equation}
  \label{e:coeff1}
  c_n(T) = \left.\frac{1}{n!} \frac{\partial^n (p/T^4)}{\partial  
  (\mu_q/T)^n}\right|_{\mu_q = 0}. 
\end{equation}
These yield a truncated result for $p(T,\mu_q)$. 

Note that computing the coefficients $c_n$, $n\geq 2$, from these 
expressions is easier on the lattice than determining the pressure
at $\mu_B=0$, $c_0(T)$, since the latter requires an integration over
$T$ and a separate lattice simulation at $T=0$. For this reason
Ref.~\cite{All05} has no results for $c_0(T)$. Since the simulations
in Ref.~\cite{All05} were done with different parameters than those 
analyzed in Fig.~\ref{fig:2} \cite{Kar1}, it is not immediately clear 
that the QPM parameters fitted to the results of Ref.~\cite{Kar1}
can also be used to describe the simulations reported in \cite{All05}.
When analyzing the lattice data of \cite{All05} we therefore refit the 
QPM parameters to the lattice results for $c_2(T)$ (see dashed line and 
squares in Fig.~\ref{fig:Gavai} below) and then assess the quality of the 
model fit by its ability to also reproduce $c_4(T)$ and $c_6(T)$ extracted 
from the same set of simulations, as well as other thermodynamic quantities 
calculated from these coefficients through Taylor expansions of the type 
(\ref{e:presdecomp}). The QPM parameters obtained by fitting $c_2(T)$ 
from \cite{All05} are \cite{Bluhm04} $\lambda = 12.0$, $T_s=0.87\,T_c$, 
and $b=426.05$ (again using $T_c=175$ MeV) \cite{fn1}.

Evaluation of the derivatives in (\ref{e:coeff1}) within the QPM is 
straightforward; for explicit analytical expressions for $c_{2,4,6}(T)$ 
we refer the reader to equations (6,\,7,\,8) in the second paper of 
Ref.~\cite{Bluhm04}. That paper also shows that the quasiparticle model
gives an excellent fit to $c_2(T)$ from \cite{All05}, and that with the
same set of parameters the QPM expressions for $c_4(T)$ and $c_6(T)$ yield 
impressive agreement with the lattice data \cite{All05}, too. In 
particular, several pronounced structures seen in $c_4(T)$ and $c_6(T)$ 
are quantitatively reproduced \cite{Bluhm04}. This constitutes a stringent 
test of the efficiency of our QPM parametrization.

\begin{figure}[t]
  \includegraphics[bb=80 10 610 710,scale=0.35,angle=-90.,clip=]%
                  {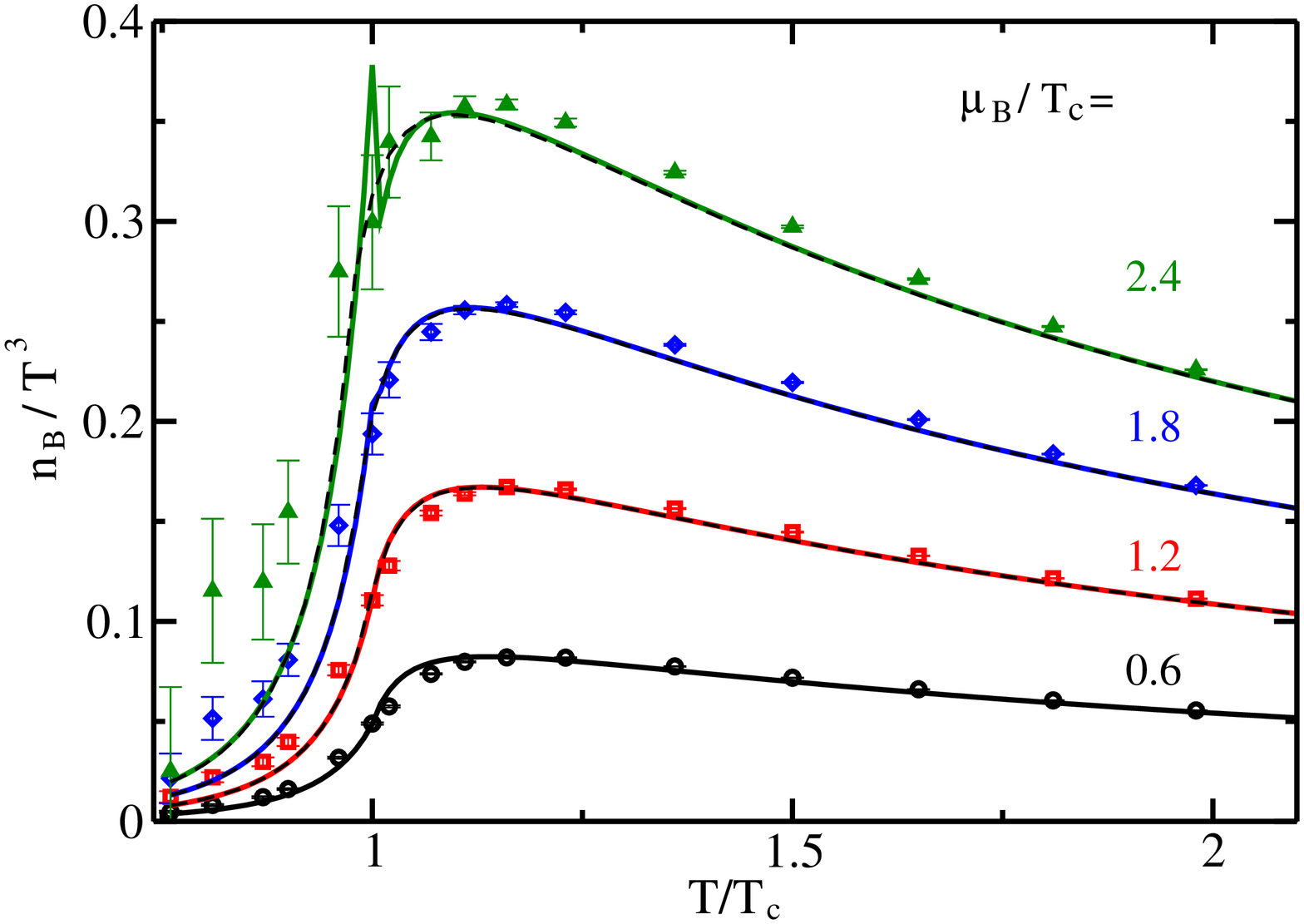}\\[-3ex]
  \includegraphics[bb=80 10 575 710,scale=0.35,angle=-90.,clip=]{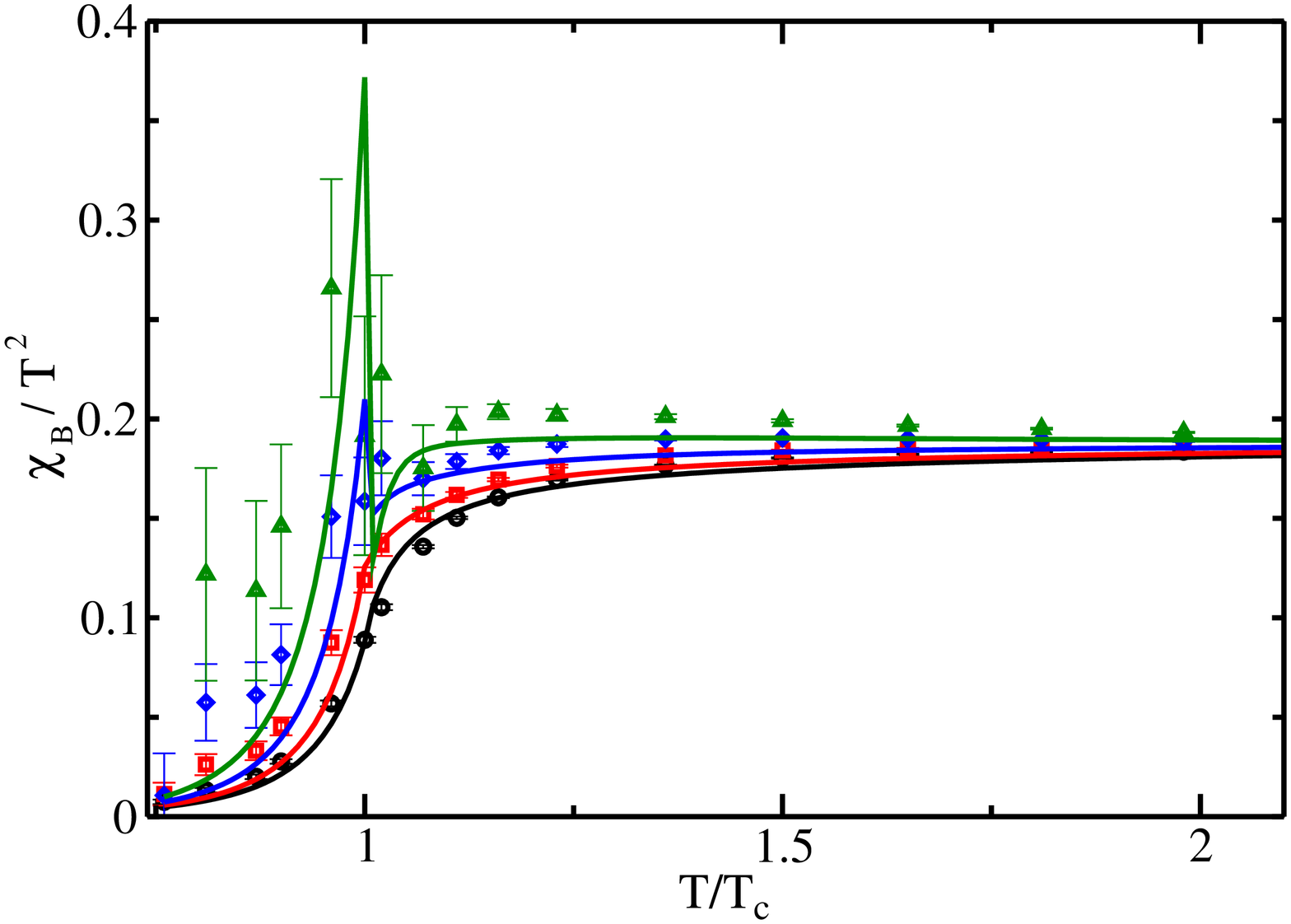}
  \caption{(Color online) Scaled baryon density $n_B/T^3$ (upper panel) and 
  baryon number susceptibility $\chi_B/T^2$ (lower panel) as a 
  function of $T/T_c$, for $\mu_B/T_c=$ 2.4, 1.8, 1.2, 0.6 (from 
  top to bottom). QPM results from the truncated expansions 
  (\ref{e:dens2}) and (\ref{e:bsuscept}) (solid lines) are compared with 
  lattice QCD data (symbols) from \cite{All05} for $N_f=2$. 
  Dashed lines in the upper panel represent the full QPM result 
  (\ref{e:dens1}) for $n_B=n_q/3$. The QPM parameters are 
  $\lambda = 12.0$, $T_s=0.87\,T_c$, and $b=426.05$, for $T_c=175$ MeV.
  \label{fig:truncs}}
\vspace*{-5mm}
\end{figure}

We here use these first three expansion coefficients $c_{2,4,6}(T)$ 
to write down truncated expansions for the net baryon density 
$n_B=\partial p/\partial\mu_B$ and the corresponding baryon number 
susceptibility $\chi_B=\partial n_B/\partial\mu_B$ which is a measure 
of fluctuations in $n_B$: 
\begin{eqnarray}
  \label{e:dens2}
  &&\frac{n_B(T,\mu_B)}{T^3} \approx 
  \\ 
  &&\quad 
  \frac{2}{3} c_2(T) \left(\frac{\mu_B}{3T}\right) + 
  \frac{4}{3} c_4(T) \left(\frac{\mu_B}{3T}\right)^3 + 
  2 c_6(T) \left(\frac{\mu_B}{3T}\right)^5 \!\!,
  \nonumber\\
  \label{e:bsuscept}
  &&\frac{\chi_B(T,\mu_B)}{T^2} \approx
  \\
  &&\quad
  \frac{2}{9} c_2(T) + 
  \frac{4}{3} c_4(T) \left(\frac{\mu_B}{3T}\right)^2 + 
  \frac{10}{3} c_6(T) \left(\frac{\mu_B}{3T}\right)^4 .
  \nonumber
\end{eqnarray}
In Fig.~\ref{fig:truncs}, the truncated QPM results for $n_B/T^3$ and 
$\chi_B/T^2$ are compared for various values of $\mu_B/T_c$ with 
lattice QCD results that were obtained from Eqs.~(\ref{e:dens2}) and 
(\ref{e:bsuscept}) with the coefficients $c_{2,4,6}(T)$ from 
\cite{All05}. We find good agreement with the lattice results; even 
below $T_c$, where our QPM parametrization is not well justified and
should be replaced by a realistic hadron resonance gas (see 
Sec.~\ref{sec:FamilyEoS}), the deviations are small but increase with 
increasing $\mu_B/T_c$. All in all, the QPM model appears to provide an 
efficient and economic parametrization of the lattice data down to 
$T\sim0.75\,T_c$.

Within the QPM model we can assess the truncation error made in 
Eqs.~(\ref{e:dens2}) by comparing this expression with the exact result
(\ref{e:dens1}) (dashed lines in the upper panel of Fig.~\ref{fig:truncs}).
The authors of \cite{All05} estimated the error induced in 
Eq.~(\ref{e:presdecomp}) by keeping only terms up to $n=4$ to remain
$\le10\%$ for $\mu_B/T\le 3$. Here we keep the terms $\sim(\mu_B/T)^6$ 
and, as the upper panel of Fig.~\ref{fig:truncs} shows, the resulting
truncated expessions for the baryon density $n_B$ agree with the exact 
results within the linewidth as long as $\mu_B/T_c\le 1.8$. For 
$\mu_B/T_c = 2.4$ we see significant deviations between the truncated 
%
%%%%%%%%%%%%%%%%%%%%%%%% Fig. 2 %%%%%%%%%%%%%%%%%%%%%%%%%%%%%%%%%%%%%%%%%%%
%%%%%%%%%%%%%%%%%%%%%%%%%%%%%%%%%%%%%%%%%%%%%%%%%%%%%%%%%%%%%%%%%%%%%%%%%%%
%
and exact expressions near $T=T_c$ which, however, can be traced back 
to an artificial mechanical instability $\partial p/\partial n_B\le 0$ 
induced by the truncation. Similar truncation effects near $T=T_c$ are 
stronger and more visible in the susceptibility $\chi_B$ (lower panel 
of Fig.~\ref{fig:truncs}). In both cases the full QPM expression is 
free of this artifact and provides a thermodynamically consistent 
description. 

We next compare the Taylor series expansion coefficients of the energy 
and entropy densities given in Ref.~\cite{Ejiri} with our model. We have 
the following decompositions \cite{Ejiri}:
\begin{equation}
\label{e:edecomp}
  e = 3 p + T^4\sum_{n=0}^\infty c_n'(T)\left(\frac{\mu_q}{T}\right)^n , 
%\\
\end{equation}
\begin{equation*}
%\label{e:sdecomp}
  s = s(T,\mu_q{=}0) + T^3\sum_{n=2}^\infty 
  \left((4{-}n)c_n(T) + c_n'(T)\right)\left(\frac{\mu_q}{T}\right)^n \!\!,
\end{equation*}
with $p$ from (\ref{e:presdecomp}), $c_n'(T)=T dc_n(T)/dT$, and 
\begin{equation}
  \label{e:s0}
  s(T,\mu_q{=}0) = T^3 \bigl(4c_0(T) + c_0'(T)\bigr) .
\end{equation}
Since these expressions contain both $c_n(T)$ and their derivatives with 
respect to $T$, $c_n'(T)$, they provide a more sensitive test of the 
model than considering the pressure alone. The expressions (\ref{e:edecomp}) 
%and (\ref{e:sdecomp}) 
can be read as Taylor series expansions with expansion coefficients 
\begin{eqnarray}
  \label{e:en}
  \frac{e}{T^4} = \sum_n e_n(T)\left(\frac{\mu_q}{T}\right)^n\!\!,&& 
  e_n(T) = 3 c_n(T) + c_n'(T),\ \ 
\\
%  \label{e:sn}
  \frac{s}{T^3} = \sum_n s_n(T)\left(\frac{\mu_q}{T}\right)^n\!\!,&&  
  s_n(T) = (4{-}n) c_n(T) + c_n'(T).
\nonumber
\end{eqnarray}
%
%%%%%%%%%%%%%%%%%%%%%%%%%%% Fig. 3 %%%%%%%%%%%%%%%%%%%%%%%%%%%%%%%%%%%%%%%%%%%
\begin{figure}[ht] 
  \includegraphics[bb=80 40 590 710,scale=0.34,angle=-90.,clip=]{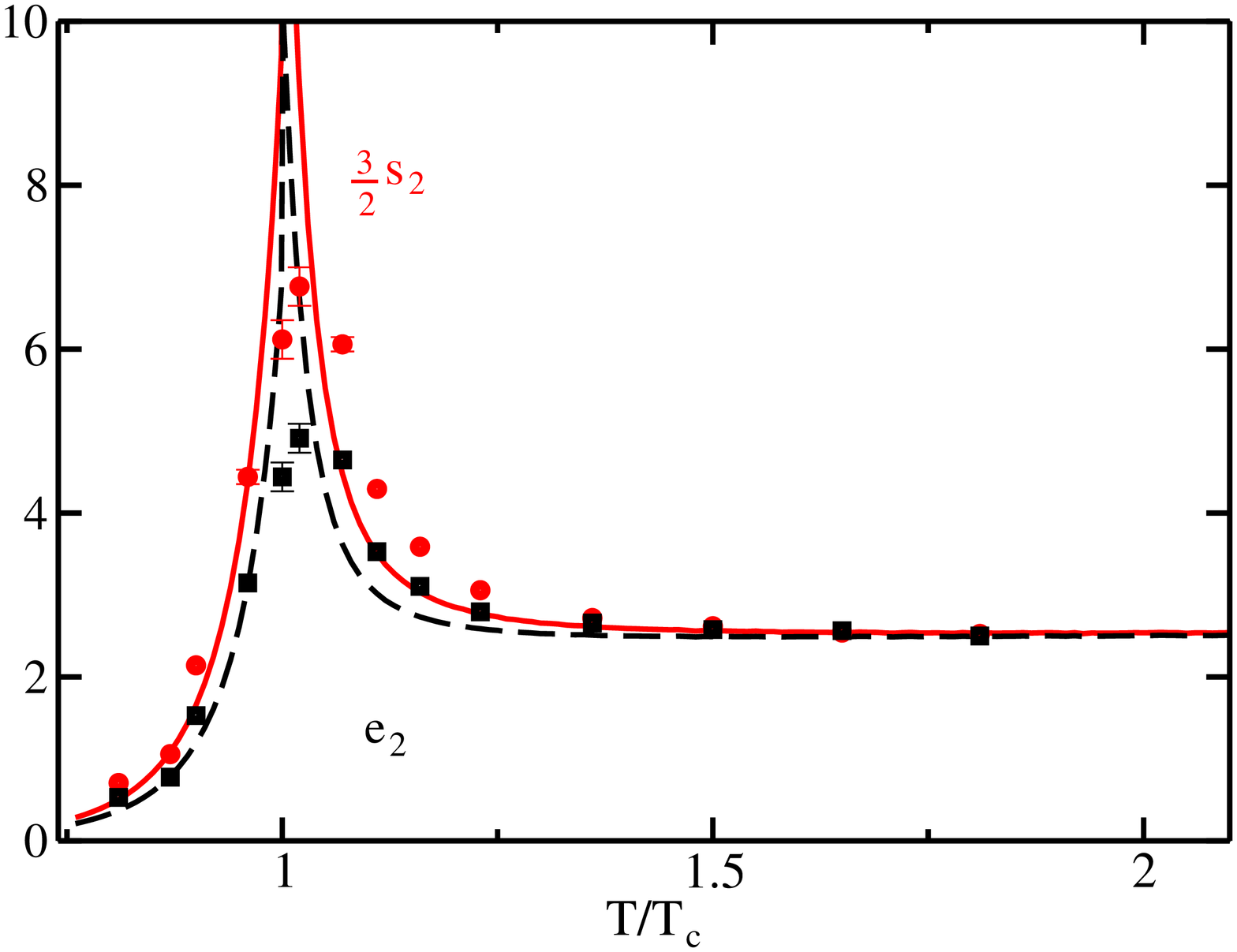}
  \includegraphics[bb=80 40 590 710,scale=0.34,angle=-90.,clip=]{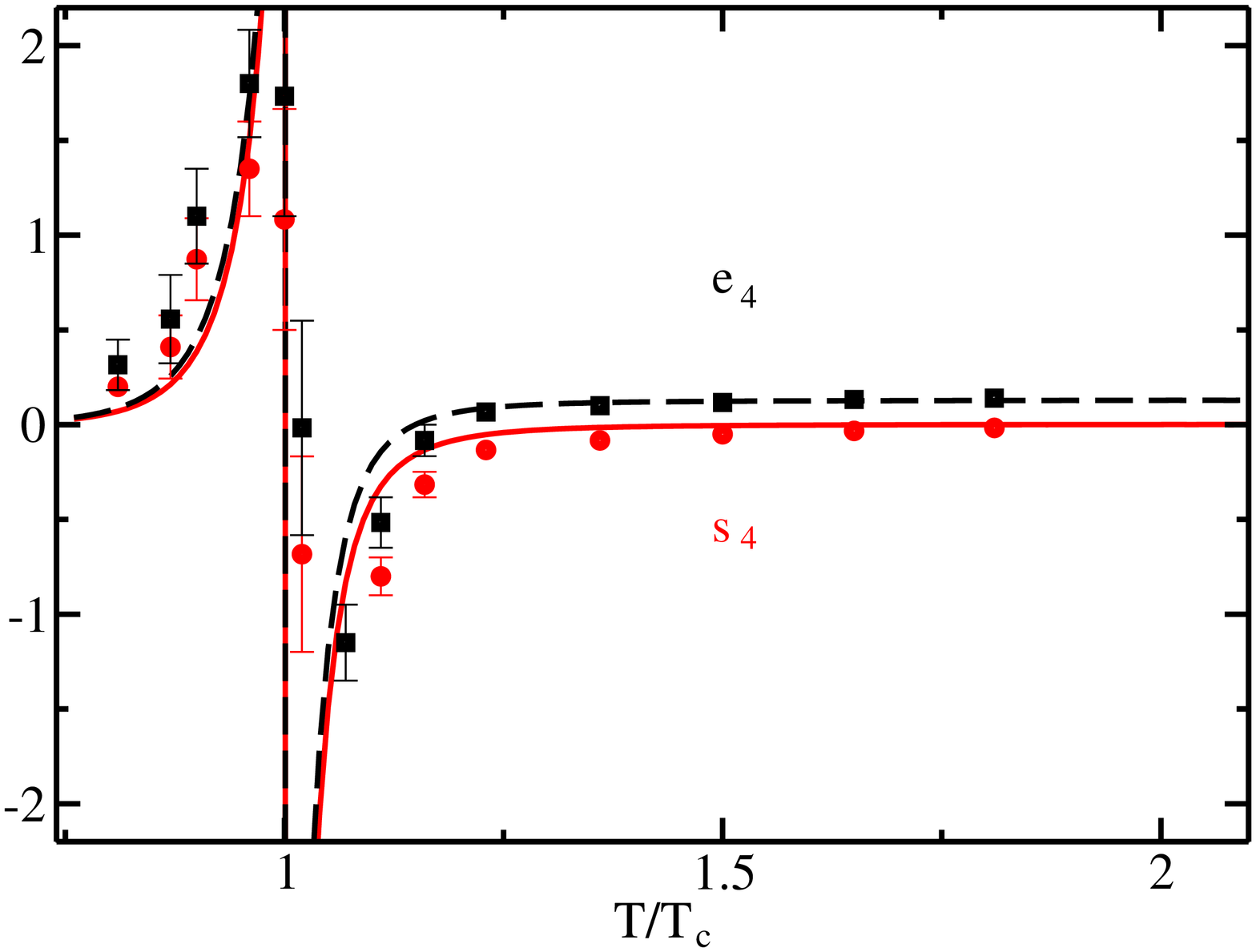}
  \caption{(Color online) Comparison of the Taylor series expansion coefficients for $e_n(T)$ 
  (squares/dashed black lines) and $s_n(T)$ (circles/solid red lines) for 
  $N_f=2$ from \cite{Ejiri} with the QPM (same parameters as in 
  Fig.~\ref{fig:truncs}). [Upper panel: $n=2$. Lower panel: $n=4$.] For 
  details see text.
  \label{fig:coeffs2}}
\end{figure}
%%%%%%%%%%%%%%%%%%%%%%%%%%%%%%%%%%%%%%%%%%%%%%%%%%%%%%%%%%%%%%%%%%%%%%%%%%%%%%
%
Figure \ref{fig:coeffs2} shows a comparison of the QPM results for 
$e_{2,4}$ and $s_{2,4}$ (obtained through fine but finite difference 
approximations of the $c_n(T)$) with the corresponding lattice QCD results 
from Ref.~\cite{Ejiri}. The QPM parameters are the same as in 
Fig.~\ref{fig:truncs}, and the agreement with the lattice data is fairly 
good. The pronounced structures observed in the vicinity of the transition 
temperature are a result of the change in curvature of $G^2(T,\mu_q{=}0)$ 
at $T{\,=\,}T_c$ (see Eq.~(\ref{e:G2param})). Note that the derivatives 
$c_n'(T)$ were estimated in \cite{Ejiri} by finite difference 
approximations of the available lattice QCD results for $c_n(T)$. 
After adjusting the difference approximation in our QPM to the lattice 
procedure, the pronounced structures in the vicinity of $T_c$ are much 
better reproduced \cite{Bluhmnew}.

We close this subsection with a calculation of the quark number 
susceptibilities which play a role in the calculation of event-by-event 
fluctuations of conserved quantities such as net baryon number, isospin 
or electric charge \cite{Asa00a,Jeo00a,Koda1,Koda2}. Across the 
quark-hadron phase transition they are expected to become large. 
For instance, the peak structure in $c_4(T)$ (which for small $\mu_B/T$
gives the dominant $\mu_B$-dependence of $\chi_B$, see Eq.~(\ref{e:bsuscept}))
can be interpreted as an indication for critical behavior. Quark number 
susceptibilities have been evaluated in lattice QCD simulations by
Gavai and Gupta \cite{Gav1}, using constant bare quark masses 
$m_{q0}=0.1\,T_c$ with $T_c$ fixed by $m_\rho/T_c=5.4$. Introducing
separate chemical potentials for $u$ and $d$ quarks and considering a 
simultaneous expansion of the QCD partition function $Z(T,\mu_u,\mu_d)$ 
in terms of $\mu_u$ and $\mu_d$, the leading $\mu_{u,d}$-independent 
contribution to the quark number susceptibility $\chi_q=9\chi_B$ can be 
expressed in terms of $\chi_{uu}$, $\chi_{ud}$ and $\chi_{dd}$ where
\begin{equation}
  \label{e:Gavai}
  \chi_{ab}=\left.\frac{\partial^2 p(T,\mu_u,\mu_d)}
                       {\partial\mu_a\partial\mu_b}\right|_{\mu_a=\mu_b=0} .
\end{equation}
These linear quark number susceptibilities can be related to the Taylor 
series expansions in (\ref{e:presdecomp}) and~(\ref{e:bsuscept}) through 
\begin{equation}
  \label{e:connection}
  c_2(T) = \frac{1}{2T^2}\left(\chi_{uu}+2\chi_{ud}+\chi_{dd}\right). 
\end{equation}
For $m_u=m_d$ one finds $\chi_{uu}=\chi_{dd}$. In Fig.~\ref{fig:Gavai}
we compare lattice QCD results \cite{Gav1} for 
$(\chi_{uu}+\chi_{ud})/T^2\equiv c_2(T)$ with a QPM fit. The QPM
parameters are adjusted to the lattice data from \cite{Gav1}, after
extrapolating the latter to the continuum by multiplying with a
factor $d=0.465$ as advocated in \cite{Gav4}. For comparison, we 
also show $c_2(T)$ from \cite{All05} and the corresponding QPM 
parametrization from Fig.~\ref{fig:truncs}. Note that the latter 
data have not yet been extrapolated to the continuum. 
%\begin{turnpage}
%
%%%%%%%%%%%%%%%%%%%%%%%%%%%%%%%%%% Fig. 4 %%%%%%%%%%%%%%%%%%%%%%%%%%%%%%%%%%%%
\begin{figure}[ht]
  \includegraphics[bb=80 10 580 710,scale=0.33,angle=-90.,clip=]{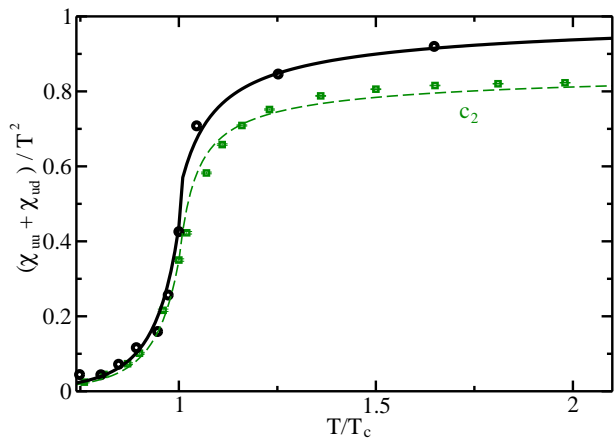}
  \caption{(Color online) Comparison of the QPM result for $(\chi_{uu}+\chi_{ud})/T^2$ 
    (solid line) with lattice QCD data (circles) from~\cite{Gav1} for 
    $N_f=2$, extrapolated to the continuum as suggested in~\cite{Gav4}. 
    The QPM parameters are $\lambda=7.0$, $T_s=0.76\,T_c$, and $b=431$,
    with $T_c=175$ MeV. For comparison, we also show lattice QCD data for
    $c_2(T)$ for $N_f=2$ from~\cite{All05} (squares) together with the
    the corresponding QPM fit (dashed line), using the same parameters as 
    in Fig.~\ref{fig:truncs}.
    \label{fig:Gavai}}
\end{figure}
%%%%%%%%%%%%%%%%%%%%%%%%%%%%%%%%%%%%%%%%%%%%%%%%%%%%%%%%%%%%%%%%%%%%%%%%%%%%%%
%
%\end{turnpage}
If we performed a continuum extrapolation of the $c_2(T)$ data from 
\cite{All05} by a factor $d=1.1$ for $T\ge T_c$ as in the case of $c_0(T)$ 
(cf. Fig.~\ref{fig:2}), 
both results would agree at large $T$ within 1\%. In the transition region
some deviations would remain, due to the different bare quark masses 
and actions employed in Refs.~\cite{All05} and \cite{Gav1}.

%%%%%%%%%%%%%%%%%%%%%%%%%%%%%%%%%%%%%%%%%%%%%%%%%%%%%%%%%%%%%%%%%%%%%%%%%%%%%
\subsection{Isentropic trajectories for $N_f=2$ quark flavors}
\label{sec:Trajectories}
%%%%%%%%%%%%%%%%%%%%%%%%%%%%%%%%%%%%%%%%%%%%%%%%%%%%%%%%%%%%%%%%%%%%%%%%%%%%%

Ideal relativistic hydrodynamics \cite{KSH,TLS,Shu,Huovinen,Huov05,Heinz_SQM04}
is considered to be the appropriate framework for describing the expansion 
of strongly interacting quark-gluon matter created in relativistic heavy-ion 
collisions. This approach requires approximate local thermal equilibrium and 
small dissipative effects. Since the fireballs created in heavy-ion 
experiments are small, pressure gradients are big and expansion rates 
are large, thermalization must be maintained by sufficiently fast momentum 
transfer rates resulting in microscopic thermalization time scales which are 
short compared to the macroscopic expansion time. The hydrodynamic description 
remains valid as long as the particles'\ mean free paths are much smaller 
than both the geometric size of the expanding fireball and its Hubble radius.

The hydrodynamic equations of motion result from the local conservation 
laws for energy-momentum and conserved charges, $\partial_\mu T^{\mu\nu}
(x){\,=\,}0$ and $\partial_\mu j_i^\mu (x){\,=\,}0$. Here, $T^{\mu\nu}$ 
denotes the energy-momentum stress tensor and $j_i^\mu$ the four-current 
of conserved charge $i$ at space-time coordinate $x$. Heavy-ion collisions
are controlled by the strong interaction which conserves baryon number, 
isospin, and strangeness. If we assume zero net isospin and strangeness
densities in the initial state, only the conservation of the baryon
number four-current $j_B^\mu$ needs to be taken into account dynamically.

The ideal fluid equations are obtained by assuming locally thermalized 
momentum distributions in which case $T^{\mu\nu}$ and $j_B^\mu$ take on 
the simple ideal fluid forms $T^{\mu\nu} = (e+p)u^\mu u^\nu -p g^{\mu\nu}$ 
and $j_B^\mu = n_B u^\mu$ \cite{Landau}. Here $g^{\mu\nu}$ is the Minkowski 
metric, $u^\mu(x)$ the local four-velocity of the fluid, and $e(x)$, $p(x)$ 
and $n_B(x)$ denote the energy density, pressure, and baryon density in the 
local fluid rest frame. The resulting set of 5 equations of motion for 6 
unknown functions is closed by the EoS which relates $p,\,e,\,$ and $n_B$. 
This is where the lattice QCD data and our QPM parametrization of the 
lattice EoS enter the description of heavy-ion collision dynamics.

Once the initial conditions are specified, the further dynamical evolution
of the collision fireball is entirely controlled by this EoS. Specifically,
the accelerating power of the fluid (i.e. its reaction to pressure gradients
which provide the thermodynamic force driving the expansion) is entirely 
controlled by the (temperature dependent) speed of sound, 
$c_s=\sqrt{\partial p/\partial e}$. To the extent that ideal fluid dynamics
is a valid description and/or dissipative effects can be controlled, the 
observation of collective flow patterns in heavy-ion collisions can thus 
provide constraints on the EoS of the matter formed in these collisions.

Ideal fluid dynamics is entropy conserving, i.e. the specific entropy
$\sigma\equiv s/n_B$ of each fluid cell (where $s$ is the local entropy 
density) stays constant in its comoving frame. Although different cells 
usually start out with different initial specific entropies, and thus
the expanding fireball as a whole maps out a broad band of widely varying 
$s/n_B$ values, each fluid cell follows a single line of constant $s/n_B$ 
in the $T{-}\mu_B$ phase diagram. It is therefore of interest to study the 
characteristics of such isentropic expansion trajectories, in particular
the behavior of $p/e$ or $c_s^2=\frac{\partial p}{\partial e}$ along them.

The isentropic trajectories for different values of $s/n_B$ follow 
directly from the first principles evaluation of the lattice EoS and
its QPM parametrization considered in the previous subsection. For 
$N_f=2$ dynamical quark flavors, the truncated Taylor series expansions 
for baryon number and entropy density with expansion coefficients $c_n(T)$ 
and $s_n(T)$ according to (\ref{e:en}) were employed  in Ref. \cite{Ejiri}
to determine the isentropic trajectories for $s/n_B=$ 300, 45, 30, 
sampling those regions of the phase diagram which can be explored with 
heavy-ion collisions at RHIC, SPS, and AGS/SIS300, respectively. In order 
to directly compare the QPM with these lattice results, we calculate 
$n_B$ from (\ref{e:dens2}) and $s$ from (\ref{e:edecomp},~\ref{e:s0}) up 
to $\mathcal{O}((\mu_B/T)^6)$, where $c_{2,4,6}(T)$ are obtained from 
(\ref{e:coeff1}), $c_0(T)=p(T,\mu_B=0)/T^4$ from (\ref{e:pres0},\ref{e:pres1}),
and the derivatives $c_n'(T)$ are estimated through fine but finite 
difference approximations of the $c_n(T)$. 

\begin{figure}[ht]
  \includegraphics[bb=80 10 580 710,scale=0.35,angle=-90.,clip=]%
                  {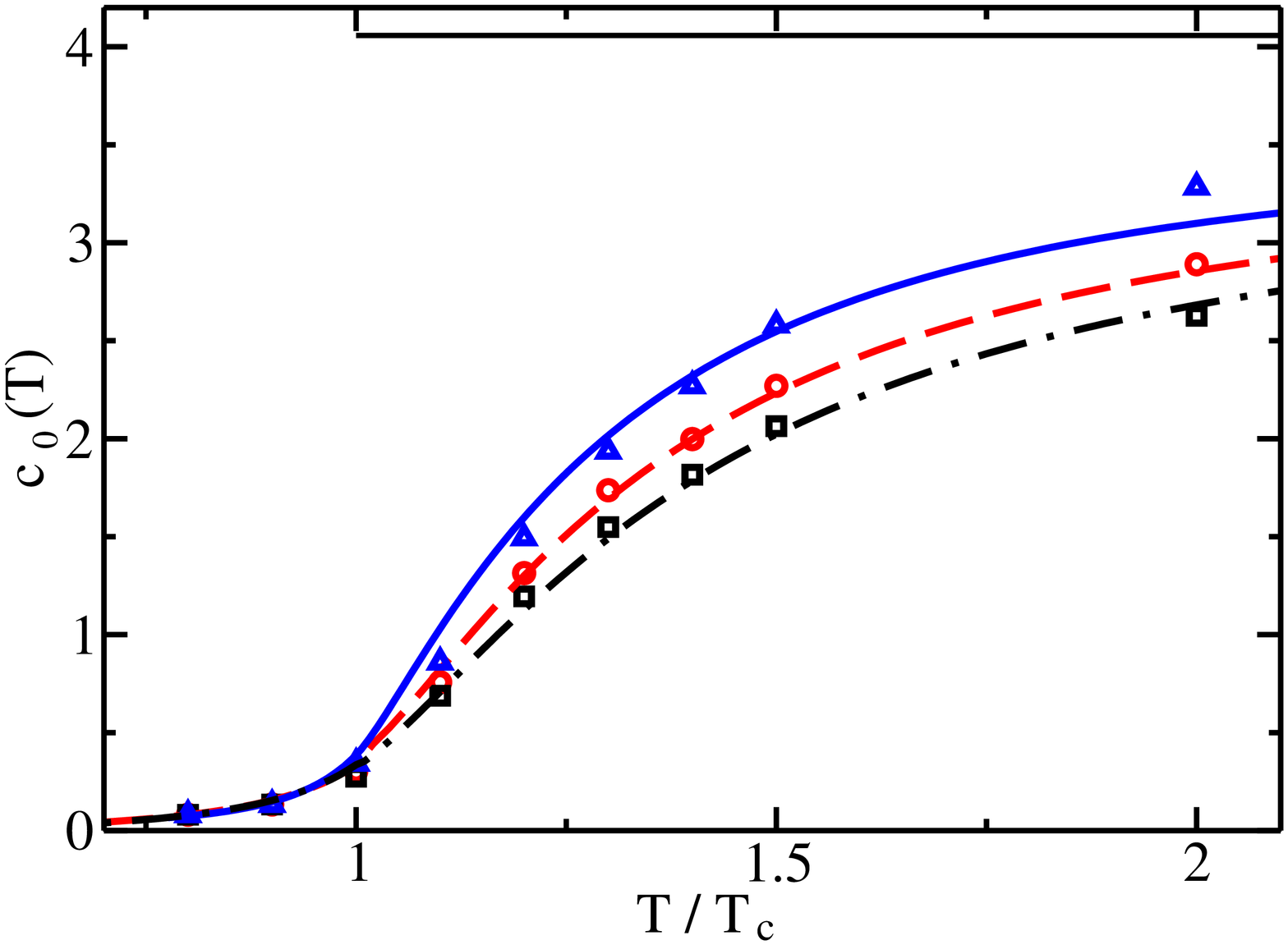}
  \includegraphics[bb=80 10 580 710,scale=0.35,angle=-90.,clip=]{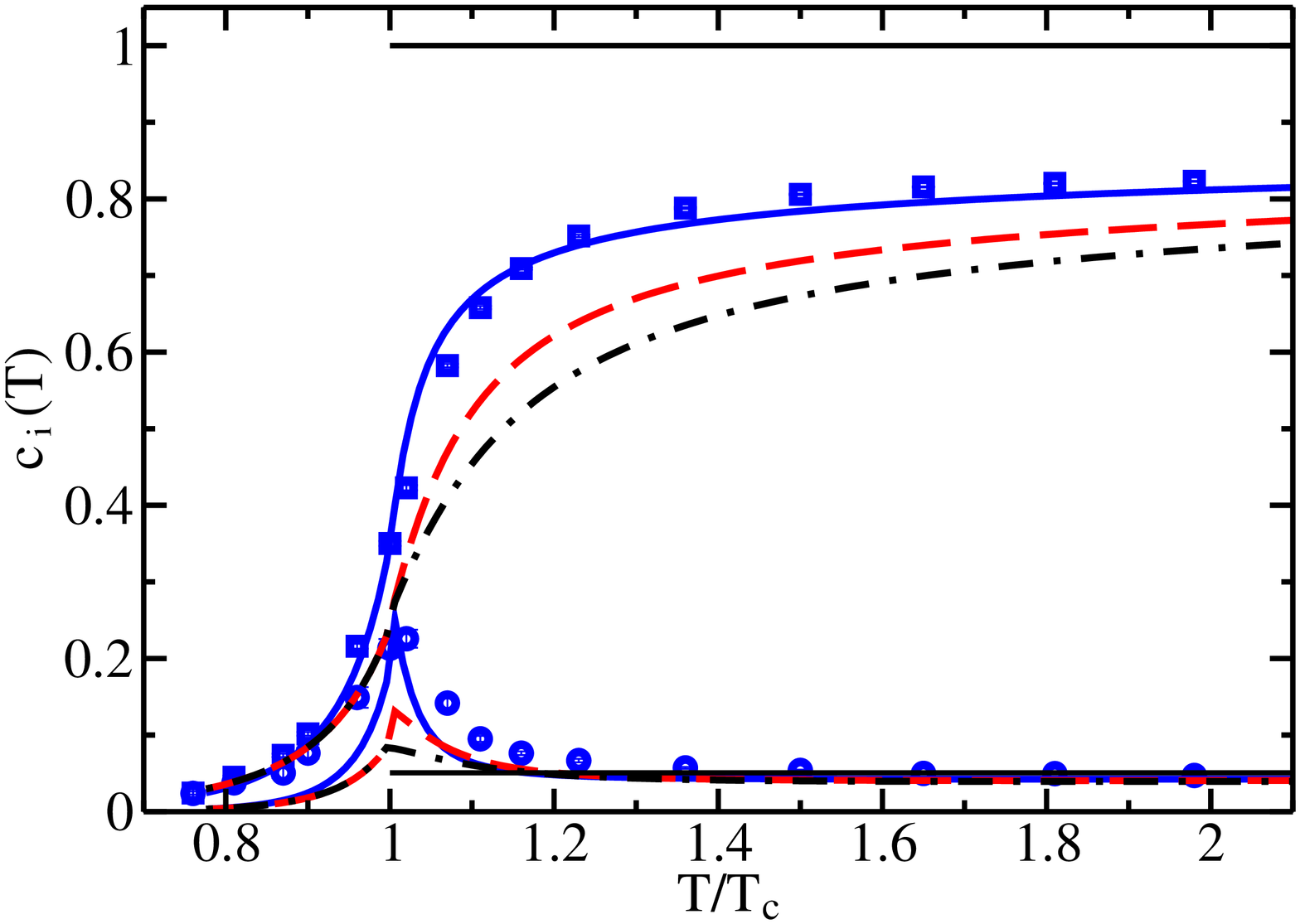}
  \caption{(Color online)
    {\sl Top panel:} $c_0(T)=p(T,\mu_B{=}0)/T^4$ as a function of $T/T_c$ 
    for $N_f=2$. Raw lattice QCD data from \cite{Kar1} (squares) and 
    guesses for the continuum extrapolated data obtained by multiplying 
    (for $T\ge T_c=175$\,MeV) by a factor $d=1.1$ (circles) and $d=1.25$ 
    (triangles) \cite{Kar1,Kar3} are shown together with the corresponding
    QPM fits (dashed-dotted, dashed, and solid curves, respectively). The 
    QPM parameters read $B(T_c)=0.31\,T_c^4$, $b=344.4$, $\lambda=2.7$, and 
    $T_s=0.46T_c$ for the dashed-dotted line (``fit 1''); they are the same 
    as in Fig.~\ref{fig:2} for the dashed line (``fit 2''); and the same as 
    in Fig.~\ref{fig:truncs} (with $B(T_c)=0.61\,T_c^4$) for the solid line 
    (``fit 3''). {\sl Bottom panel:} Corresponding QPM results compared with lattice 
    results for $c_2(T)$ (squares) and $c_4(T)$ (circles) as a function of 
    $T/T_c$ with the same line code as in the top panel. 
    The horizontal lines indicate the Stefan-Boltzmann values. 
    \label{fig:c0para}}
\end{figure}

Besides investigating the impact of different continuum extrapolations 
of $c_0(T)$ on the pattern of isentropic trajectories, we can ask whether 
the differences observed between the parametrizations of $c_0(T)$ and 
$c_2(T)$ can be absorbed in such an extrapolation. Note that, even though 
the cutoff dependence of the lattice results is qualitatively similar 
at $\mu_B=0$ and at $\mu_B\ne 0$, no uniform continuum extrapolation 
is expected for the different Taylor expansion coefficients 
\cite{All05,Karpriv}. In Fig.~\ref{fig:c0para} we show the raw lattice 
data for $c_0(T)$ \cite{Kar1} (squares) together with a continuum 
extrapolation (circles) obtained by multiplying the raw data for $T\ge T_c$ 
by a factor $d=1.1$. The corresponding QPM parametrizations (``fit 1''
(dash-dotted) and ``fit 2'' (dashed) in the upper panel of 
Fig.~\ref{fig:c0para}) reproduce the lattice QCD results impressively 
well. Nonetheless, the corresponding QPM results for $c_{2,4}(T)$ 
underpredict the lattice data, as depicted in the bottom panel of 
Fig.~\ref{fig:c0para}. In particular, the pronounced structure in 
$c_4(T)$ at $T_c$ is not well reproduced by the QPM fit. If we instead use 
a QPM parametrization that optimally reproduces $c_2(T)$ (solid line in 
the bottom panel of Fig.~\ref{fig:c0para}), the corresponding QPM result 
for $c_0(T)$ (``fit 3'' in the upper panel of Fig.~\ref{fig:c0para}) agrees
fairly well with an assumed continuum extrapolation of the raw lattice data 
by a factor $d=1.25$ for $T\ge T_c$ (triangles). 
%
%%%%%%%%%%%%%%%%%%%%%% Fig. 5 %%%%%%%%%%%%%%%%%%%%%%%%%%%%%%%%%%%%%%%%%%%%%%%%
%%%%%%%%%%%%%%%%%%%%%%%%%%%%%%%%%%%%%%%%%%%%%%%%%%%%%%%%%%%%%%%%%%%%%%%%%%%%%%
%

\begin{figure}[t]
  \includegraphics[bb=80 10 580 710,scale=0.33,angle=-90.,clip=]{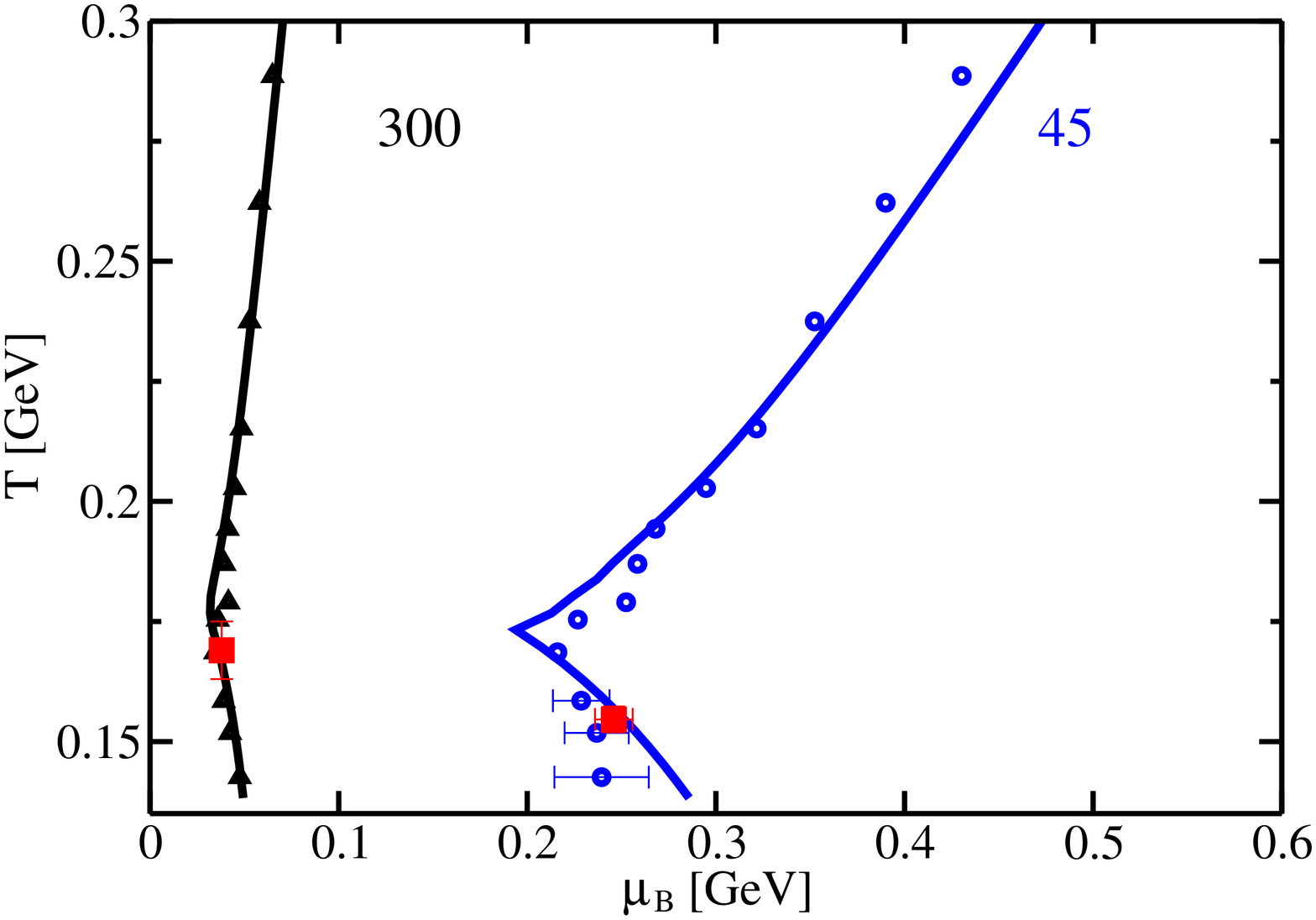}
  \includegraphics[bb=80 10 580 710,scale=0.33,angle=-90.,clip=]%
                  {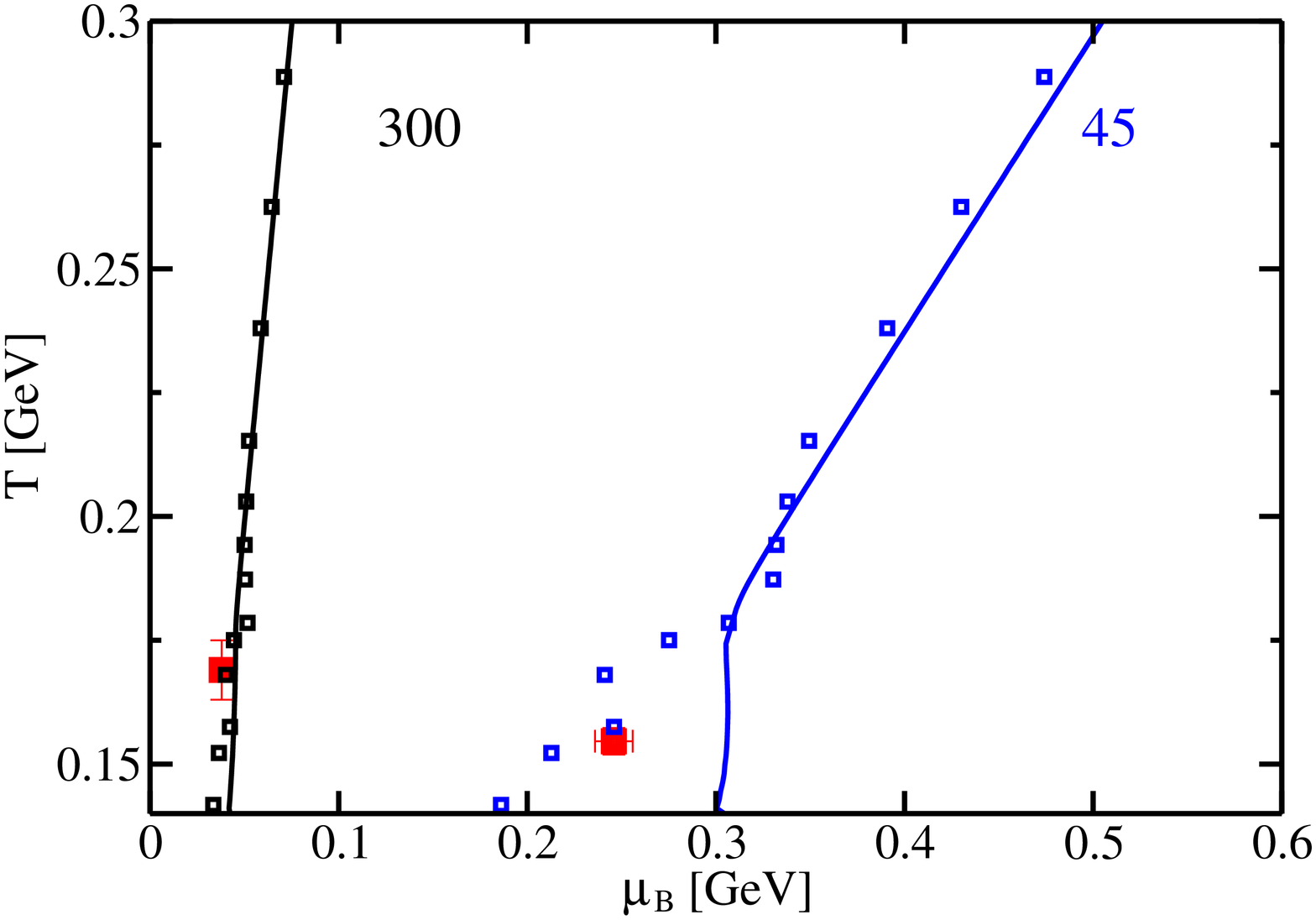}
  \caption{(Color online)
    Isentropic evolutionary paths. Triangles and circles indicate $N_f=2$ 
    lattice QCD data from~\cite{Ejiri} for $s/n_B = $ 300 and 45, 
    respectively. Corresponding QPM results are depicted in the upper panel
    for a mixed fit where $c_0(T)$ and $c_2(T)$ were fitted independently 
    (cf. Figs.~\ref{fig:2} and~\ref{fig:truncs}). In the lower panel we show
    results from ``fit 3'' from Fig.~\ref{fig:c0para}, with open squares 
    indicating the corresponding continuum-extrapolated lattice results 
    where the raw $c_0(T)$ lattice data were multiplied by a constant 
    factor $d=1.25$ at $T\ge T_c$ \cite{Kar1}. Full red squares show 
    chemical freeze-out points deduced in \cite{Cley05,Mann06} from 
    hadron multiplicity data, as summarized in~\cite{Cley06}. 
\label{fig:isentrops}}
\end{figure}

In Fig.~\ref{fig:isentrops}, the QPM results for $s/n_B=$ 300 and 45 
employing different fits are exhibited together with the results of 
\cite{Ejiri}. In the top panel of Fig.~\ref{fig:isentrops} we see that the 
lattice results can be fairly well reproduced when using simultaneously two 
separately optimized QPM parametrizations for $c_0(T)$ and $c_2(T)$ 
(cf. Fig.~\ref{fig:2} and~\ref{fig:truncs}). This approach, however, would give 
up thermodynamic consistency of the model. When using a single consistent 
parametrization for both $c_0$ and $c_2$, specifically the one shown by 
the solid lines in Fig.~\ref{fig:c0para} corresponding to ``fit 3'', the 
QPM produces the isentropes shown in the bottom panel of 
Fig.~\ref{fig:isentrops}. (The other two fits shown in Fig.~\ref{fig:c0para}
%
%%%%%%%%%%%%%%%%%%%%%%%%% Fig. 6 %%%%%%%%%%%%%%%%%%%%%%%%%%%%%%%%%%%%%%%%%%%%%
%%%%%%%%%%%%%%%%%%%%%%%%%%%%%%%%%%%%%%%%%%%%%%%%%%%%%%%%%%%%%%%%%%%%%%%%%%%%%%
%
yield almost the same isentropic expansion trajectories as ``fit 3''.) 
For large $s/n_B$, i.~e. for small net baryon densities, differences 
between the QPM results in the top and bottom panels of 
Fig.~\ref{fig:isentrops} are small, although the top fit shows a weak 
structure near $T_c$ which disappears in the selfconsistent fit shown in the
bottom panel. With decreasing $s/n_B$ the differences between the results
from the two fitting strategies increase. They are mainly caused by 
differences in the slope of $c_0(T)$ which affect the shape of $s(T)/T^3$ 
and translate, for a given isentropic trajectory, into large variations 
of $\mu_B$ near $T_c(\mu_B{=}0)=175$ MeV while causing only small 
differences of about 6\% at large $T$. In particular, the pronounced 
structures of the isentropic trajectory near the estimated phase border 
are completely lost in the selfconsistent fit procedure. This shows that
the pattern of the isentropic expansion trajectories is quite sensitive 
to details of the EoS. For instance, when employing $c_0(T)$ data which
were extrapolated to the continuum by multiplication with a factor $d=1.25$
at $T\ge T_c$ while leaving $c_{2,4,6}(T)$ unchanged, one obtains the 
isentropic expansion trajectories shown by open squares in the bottom
panel of Fig.~\ref{fig:isentrops} which also lack any structure near the
phase transition. 

Changing the deconfinement transition temperature to $T_c=170$ MeV
results in a shift of the trajectories by about 10\% in $\mu_B$ 
direction near $T_c$ but has negligible consequences for $T\ge 1.5 \,T_c$. 
At asymptotically large $T$, where $c_{0,2}(T)$ are essentially flat, the 
relation $\frac{\mu_B}{T} = 18 \frac{c_0}{c_2}(\frac{n_B}{s})$ holds 
for small $\mu_B$, i.~e. lines of constant specific entropy are 
essentially given by lines of constant $\mu_B/T$, as is the case in
a quark-gluon plasma with perturbatively weak interactions. 

Figure~\ref{fig:isentrops} also shows the chemical freeze-out points 
deduced from hadron multiplicity data for Au+Au collisions at 
$\sqrt{s}=130\,A$\,GeV at RHIC ($T_\mathrm{chem}=169\pm 6$ MeV and 
$\mu_{B,\mathrm{chem}}=38\pm 4$ MeV \cite{Cley05}) and for 158\,$A$\,GeV
Pb+Pb collisions at the CERN SPS ($T_\mathrm{chem}=154.6\pm 2.7$ MeV and 
$\mu_{B,\mathrm{chem}}=245.9\pm 10.0$ MeV \cite{Mann06}). Note that
the specific entropies at these freeze-out points as deduced from the 
statistical model \cite{Wheaton} are $s/n_B=200$ for RHIC-130 and 
$s/n_B=30$ for SPS-158, i.e. only about 2/3 of the values corresponding 
to the QPM fit of the QCD lattice data. One should remember, though, that 
the phenomenological values are deduced from experimental data using a 
complete spectrum of hadronic resonances whereas the lattice simulations 
were performed for only $N_f=2$ dynamical quark flavors with not quite
realistic quark masses. 

\begin{figure}[ht]
  \includegraphics[bb=80 10 580 710,scale=0.35,angle=-90.,clip=]%
                  {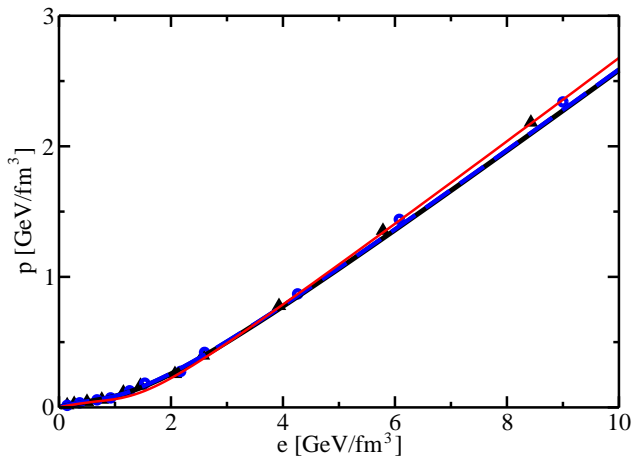}
  \caption{(Color online) 
    Lattice QCD data \cite{Ejiri} of $p$ as a function of $e$ for $N_f=2$ 
    along isentropes with $s/n_B{\,=\,}300$ (triangles) and 45 (circles), 
    compared with the corresponding QPM results (solid blue and dashed 
    black lines, respectively). These two thick lines employ the mixed 
    fit shown in the upper panel of Fig.~\ref{fig:isentrops} and are
    indistinguishable for $s/n_B{\,=\,}300$ and 45. The thin solid lines 
    show corresponding results for the self-consistent ``fit 3'' from Fig.~\ref{fig:c0para}.
    Again the curves for different $s/n_B$ are indistinuishable, and also
    the deviations from the mixed fit are minor.
    \label{fig:suppl1}}
\end{figure}

Figure~\ref{fig:suppl1} shows that along isentropic expansion lines 
the EoS is almost independent of the value of $s/n_B$. Accordingly, the 
speed of sound $c_s^2=\partial p/\partial e$ (which controls the build-up
of hydrodynamic flow) is essentially independent of the specific entropy. 
Note that whether we employ the mixed fit or the thermodynamically 
%
%%%%%%%%%%%%%%%%%%%%%%% Fig. 7 %%%%%%%%%%%%%%%%%%%%%%%%%%%%%%%%%%%%%%%%%%%%%%%
%%%%%%%%%%%%%%%%%%%%%%%%%%%%%%%%%%%%%%%%%%%%%%%%%%%%%%%%%%%%%%%%%%%%%%%%%%%%%%
%
consistent fits 1, 2 and 3 of Fig.~\ref{fig:c0para} does not significantly
affect the EoS along the isentropes; for large energy densities 
$e\gtrsim 30$\,GeV/fm$^3$ the differences in $p(e)$ are less than 2\%. 

%%%%%%%%%%%%%%%%%%%%%%%%%%%%%%%%%%%%%%%%%%%%%%%%%%%%%%%%%%%%%%%%%%%%%%%%%%%%%%
\subsection{A remark on the QCD critical point} 
\label{sec2d}
%%%%%%%%%%%%%%%%%%%%%%%%%%%%%%%%%%%%%%%%%%%%%%%%%%%%%%%%%%%%%%%%%%%%%%%%%%%%%%

At a critical point (CP) a first order phase transition line terminates 
and the transition becomes second order. QCD with $N_f=2+1$ dynamical 
quark flavors with physical masses is a theory where such a CP is expected 
at finite $T$ and $\mu_B$ \cite{Hala98,Rajagop,Steph}. Its precise location 
is still a matter of debate \cite{Gav1,Fodor,deForc03,Schmidt03}, 
but \cite{Fodor} claim $T_E=162$ MeV and $\mu_{B,E}=360$ MeV 
for the critical values. In the following, we focus on initial baryon 
densities $n_B < 0.5$ fm$^{-3}$ which, assuming isentropic expansion 
with conserved $s/n_B=250$, corresponds to a baryon chemical potential 
$\mu_B(T{=}170\,\mathrm{MeV}) < 60$ MeV. This is sufficiently far from 
the conjectured CP that we should be justified in assuming that the 
EoS is adequately parametrized by our QPM for describing bulk 
thermodynamic properties and the hydrodynamical evolution of the hot 
QCD matter.

%%%%%%%%%%%%%%%%%%%%%%%%%%%%%%%%%%%%%%%%%%%%%%%%%%%%%%%%%%%%%%%%%%%%%%%%%%%%%
\section{Equation of State 
\label{sec:FamilyEoS}}
%%%%%%%%%%%%%%%%%%%%%%%%%%%%%%%%%%%%%%%%%%%%%%%%%%%%%%%%%%%%%%%%%%%%%%%%%%%%%

In this Section we concentrate on the physical case of $N_f=2+1$ dynamical
quark flavors and match the QPM fit to the lattice QCD data at temperatures
above $T_c$ to a realistic hadron resonance gas EoS below $T_c$. In this way
we construct an EoS that can be applied to all stages of the hydrodynamic
expansion of the hot matter created in relativistic heavy-ion collisions at 
RHIC and LHC. We focus our attention on the region of small net baryon 
density explored at these colliders.

%%%%%%%%%%%%%%%%%%%%%%%%%%%%%%%%%%%%%%%%%%%%%%%%%%%%%%%%%%%%%%%%%%%%%%%%%%%%%
\subsection{Pressure as a function of energy density}
%%%%%%%%%%%%%%%%%%%%%%%%%%%%%%%%%%%%%%%%%%%%%%%%%%%%%%%%%%%%%%%%%%%%%%%%%%%%%

Our goal is to arrive at an EoS in the form $p(e,n_B)$ as needed in 
hydrodynamic applications. We anchor our QPM approach above $T_c$ 
to lattice QCD simulations for $N_f=2{+}1$ dynamical quark flavors
presented in \cite{Kar1,Peik,Kar2} where $p(T)/T^4$ and $e(T)/T^4$ were 
calculated using $m_{q0}=0.4T$ and $m_{s0}=T$. Unfortunately, Taylor 
series expansions for non-zero $\mu_B$ analogous to the $N_f=2$ case 
are not available for $N_f=2{+}1$. Effects of finite $\mu_B$ were 
studied in \cite{Fodorlat1} for $N_f=2+1$ by the multi-parameter 
reweighting method and successfully compared with the quasiparticle 
model in \cite{Szabo} by testing the extrapolation via Eq.~(\ref{e:flow}). 
We here concentrate on results from lattice QCD simulations employing 
improved actions \cite{Kar1} which strongly reduce lattice discretization 
errors at high temperatures. First, we focus on the available data at 
$\mu_B=0$ and assume that the extension to non-zero $\mu_B$ can be 
%
%%%%%%%%%%%%%%%%%%%%%%% Fig. 8 %%%%%%%%%%%%%%%%%%%%%%%%%%%%%%%%%%%%%%%%%%%%%%%
\begin{figure}[ht]
  \includegraphics[bb=80 10 580 710,scale=0.35,angle=-90.,clip=]{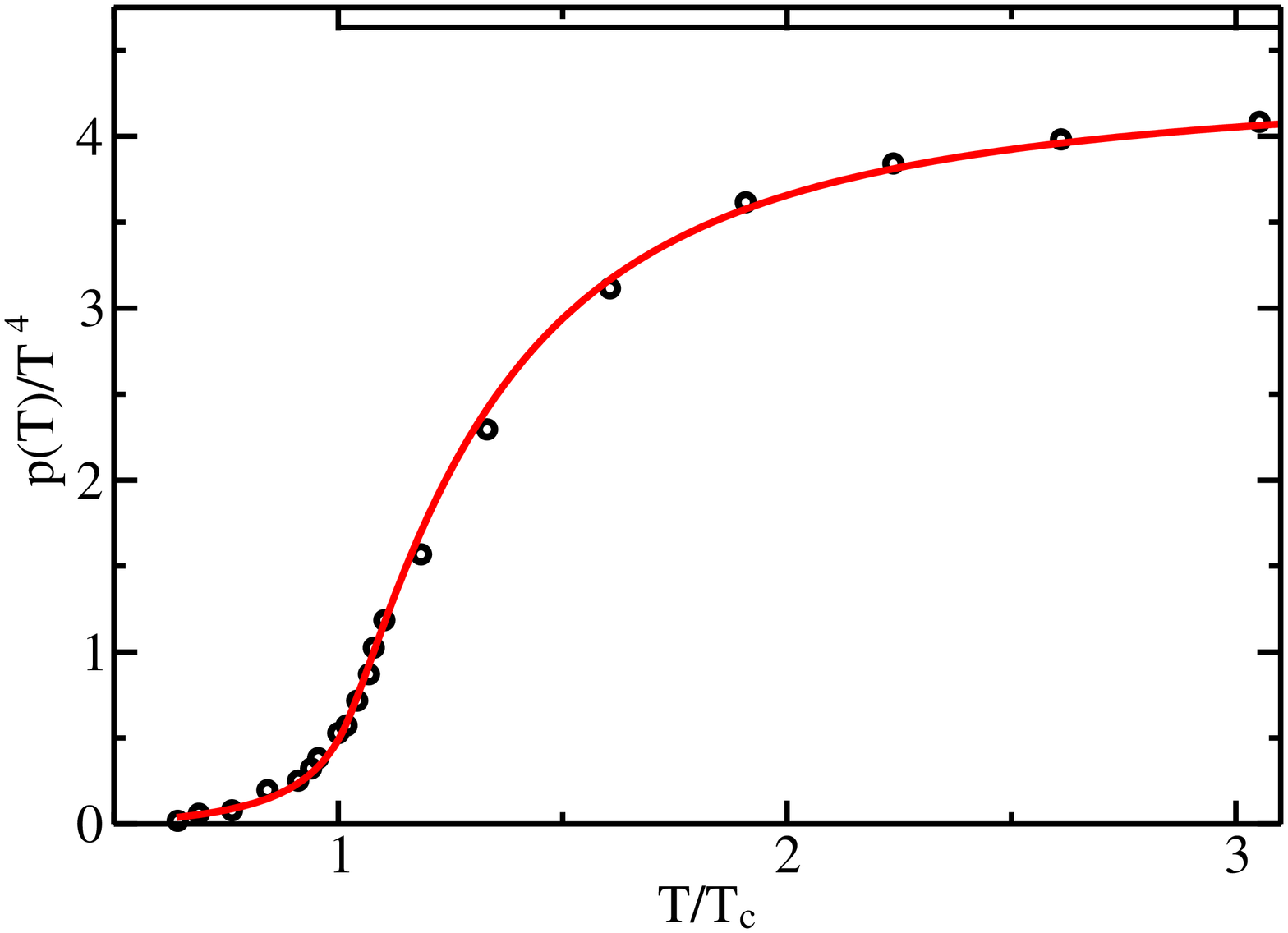}
  \includegraphics[bb=80 10 580 710,scale=0.35,angle=-90.,clip=]{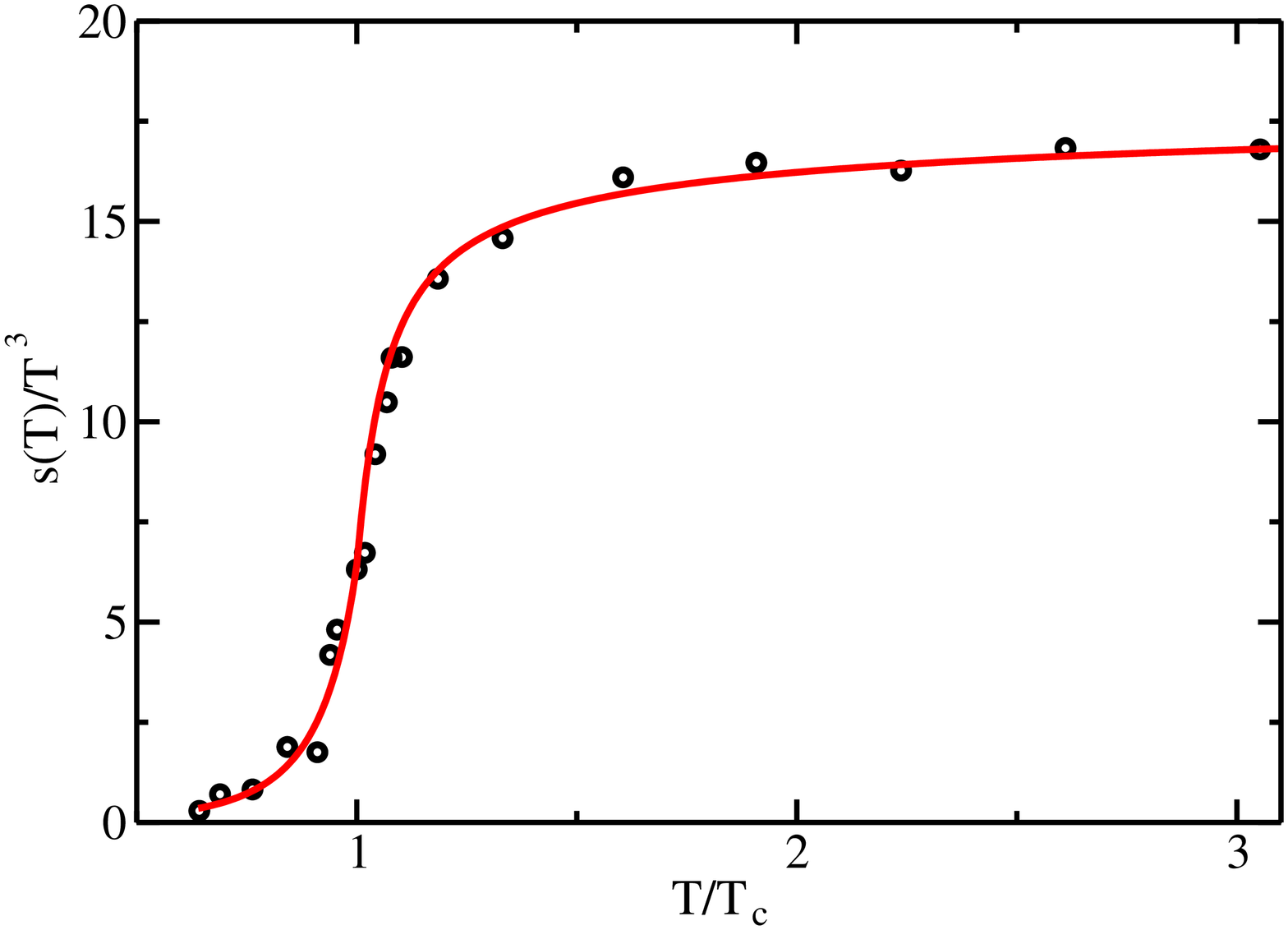}
  \caption{(Color online)
    Comparison of the QPM with lattice QCD results (symbols) for the scaled 
    pressure $p/T^4$ (top panel) and the scaled entropy density $s/T^3$ 
    (bottom panel) as a function of $T/T_c$ for $N_f=2+1$ and $\mu_B=0$. 
    The lattice QCD data~\cite{Kar2} are already continuum extrapolated. 
    The QPM parameters read $\lambda = 7.6$, $T_s=0.8T_c$, $b=348.72$ and 
    $B(T_c)=0.52 T_c^4$ where $T_c=170$ MeV. In the top panel, the
    horizontal line indicates the Stefan-Boltzmann value 
    $p_\mathrm{SB}/T^4=\bar{c}_0=(32{+}21N_f)\pi^2/180$, using $N_f{=}2.5$ 
    to account for the non-zero strange quark mass. 
\label{fig:2+1}}
\end{figure}
%%%%%%%%%%%%%%%%%%%%%%%%%%%%%%%%%%%%%%%%%%%%%%%%%%%%%%%%%%%%%%%%%%%%%%%%%%%%%%%
%
accomplished through the QPM without any complications, relying on the
successful test of our model at finite baryon density for $N_f=2$
as reported in the preceding section and earlier publications.

In Fig.~\ref{fig:2+1} we compare the QPM results for the pressure 
$p(T)/T^4$ and entropy density $s(T)/T^3$ with $N_f=2{+}1$ lattice QCD 
data where $s$ follows simply from $e$ and $p$ through $s/T^3= (e{+}p)/T^4$. 
The parametrization found at $\mu_B=0$ is now used to obtain the required 
thermodynamic observables at non-zero $n_B$ from the full QPM via
Eqs.~(\ref{e:pres0}), (\ref{e:entr0}) and the relation 
$e{+}p{-}Ts = \mu_B n_B$, exploiting the Maxwell relation (\ref{e:flow}). 

%
%%%%%%%%%%%%%%%%%%%%%%% Fig. 9 %%%%%%%%%%%%%%%%%%%%%%%%%%%%%%%%%%%%%%%%%%%%%%%
\begin{figure}[ht]
  \includegraphics[bb=80 10 580 710,scale=0.35,angle=-90.,clip=]{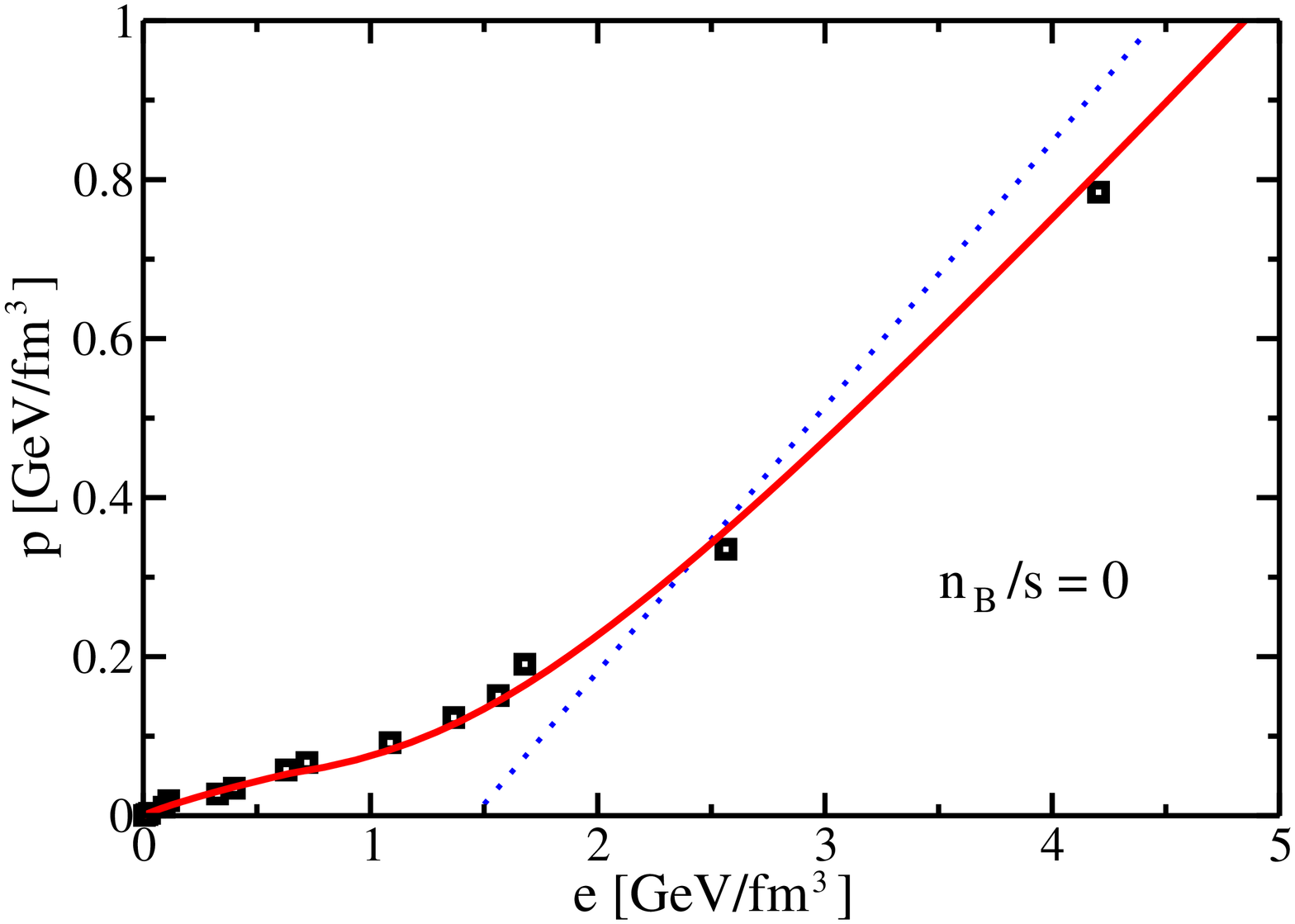}
  \includegraphics[bb=80 10 580 710,scale=0.35,angle=-90.,clip=]{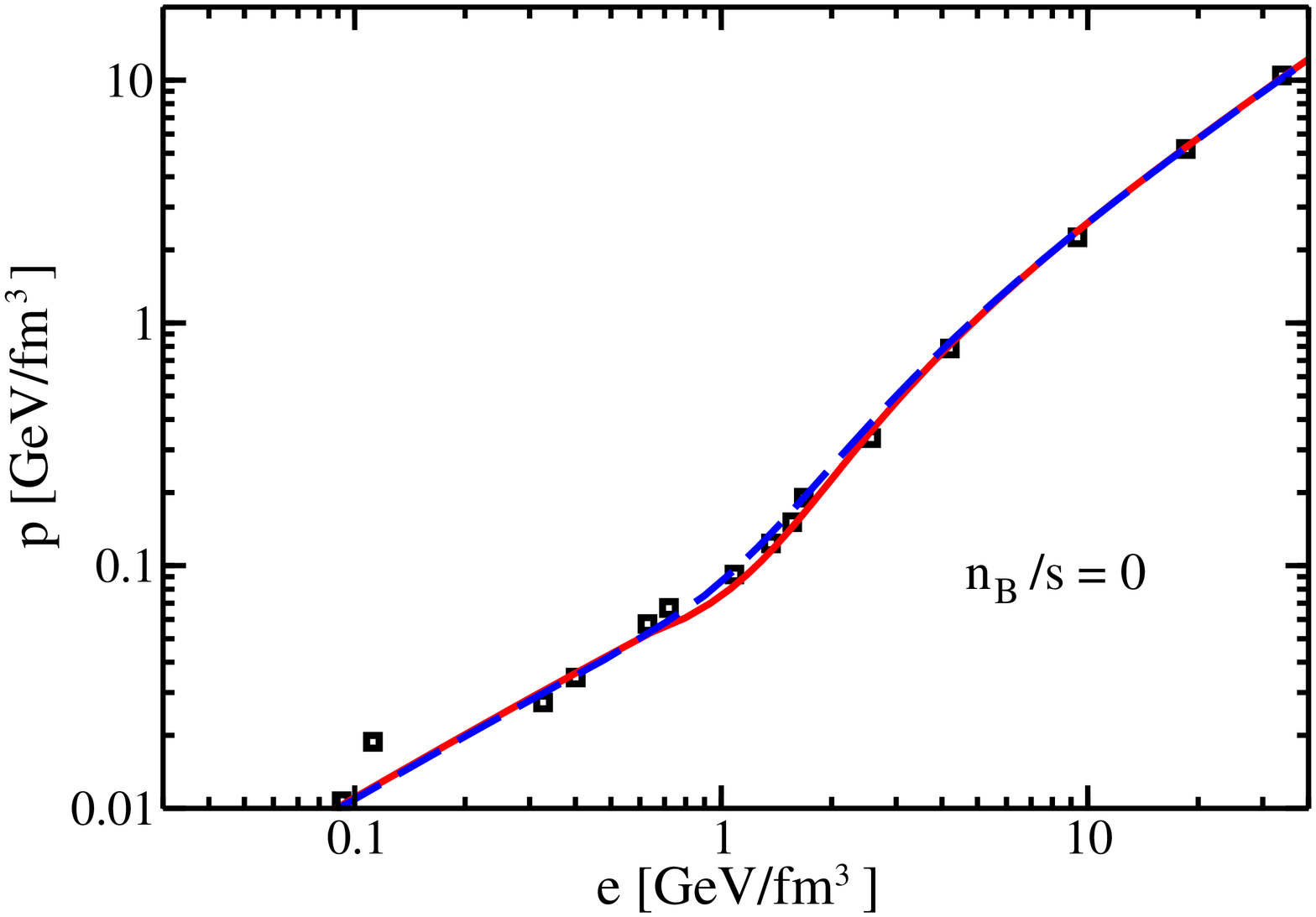}
  \caption{(Color online)
    {\sl Top panel:} $N_f{\,=\,}2{+}1$ QPM equation of state of strongly 
    interacting matter for vanishing net baryon density (solid line) 
    compared with $N_f{\,=\,}2{+}1$ continuum extrapolated lattice QCD 
    data \cite{Kar2} (squares) at $n_B{\,=\,}0$. The dotted line 
    represents $p(e)$ for a gas of massless non-interacting quarks and 
    gluons with a bag constant $B^{1/4}{\,=\,}230$\,MeV. 
    {\sl Bottom panel:} QPM EoS for $N_f{\,=\,}2$ (dashed line) employing 
    ``fit 2'' in Figs.~\ref{fig:2} and \ref{fig:c0para}, compared with 
    lattice data \cite{Kar2} (squares) and QPM results (solid line) for 
    $N_f{\,=\,}2{+}1$, in logarithmic representation.
    \label{fig:EoS}}
\end{figure}
%%%%%%%%%%%%%%%%%%%%%%%%%%%%%%%%%%%%%%%%%%%%%%%%%%%%%%%%%%%%%%%%%%%%%%%%%%%%%%%
%
In Fig.~\ref{fig:EoS} we compare the QPM equation of state $p(e)$ at 
$n_B = 0$ with the corresponding lattice QCD result deduced from data 
for $p$ and $e$ at $n_B = 0$ \cite{Kar1} in the energy density domain 
explored by heavy ion collisions at RHIC. The used lattice data \cite{Kar1} 
were already extrapolated to the continuum in \cite{Kar2}. 
In \cite{Peik,Karsch01} $T_c=(173\pm 8)$ MeV was found for the 
deconfinement transition temperature. Recent analyses \cite{Fodornew,Karsch_Tc}
have pointed out remaining uncertainties in the extraction of $T_c$ which
would have to be sorted out by simulations on larger lattices. Here, we set the 
physical scale to $T_c=170$ MeV (see discussion below). In the transition 
region the energy density $e(T)$ varies by 300\% within a temperature 
interval of $\Delta T \approx 20$\,MeV while $p(T)$ rises much more slowly
(see upper panels in Figs.~\ref{fig:2+1} and \ref{fig:EoS}). This 
indicates a rapid but smooth crossover for the phase transition from
hadronic to quark matter. At large energy densities $e\ge 30$\,GeV/fm$^{3}$
the EoS follows roughly the ideal gas relation $e=3p$. For the sake of 
comparison, a bag model equation of state describing a gas of massless 
non-interacting quarks and gluons with bag constant $B^{1/4} = 230$ MeV 
is also shown in Fig.~\ref{fig:EoS} (straight dotted line in the top 
panel).

As an aside, differences in $p(e,n_B{=}0)$ arising from considering 
different numbers $N_f$ of dynamical quark flavors are investigated 
in the bottom panel of Fig.~\ref{fig:EoS}. Comparing the QPM result 
for $N_f=2{+}1$ with the result for $N_f=2$ (see Fig.~\ref{fig:2}), 
the latter exceeds the $N_f=2{+}1$ result in the transition region 
(by about 12\% at $e=1$ GeV/fm$^3$). For larger energy densities 
$e\ge 3$ GeV/fm$^3$ the EoS is found to be fairly independent of 
$N_f$ even though at fixed $T$ both $p(T)$ and $e(T)$ are significantly 
smaller for $N_f=2$ than for $N_f=2{+}1$ (see Figs.~\ref{fig:2},
\ref{fig:2+1}). 

%%%%%%%%%%%%%%%%%%%%%%%%%%%%%%%%%%%%%%%%%%%%%%%%%%%%%%%%%%%%%%%%%%%%%%%%%%%%%
\subsection{Baryon density effects}
\label{sec3b}
%%%%%%%%%%%%%%%%%%%%%%%%%%%%%%%%%%%%%%%%%%%%%%%%%%%%%%%%%%%%%%%%%%%%%%%%%%%%%

We turn now to the baryon density dependence of the EoS. Since for 
hydrodynamics the relation $p(e,n_B)$ matters, we consider the $n_B$ 
%
%%%%%%%%%%%%%%%%%%%%%%% Fig. 10 %%%%%%%%%%%%%%%%%%%%%%%%%%%%%%%%%%%%%%%%%%%%%%%
\begin{figure}[ht]
  \includegraphics[bb=20 90 600 680,width=6.8cm,height=8cm,angle=-90.,clip=]%
                  {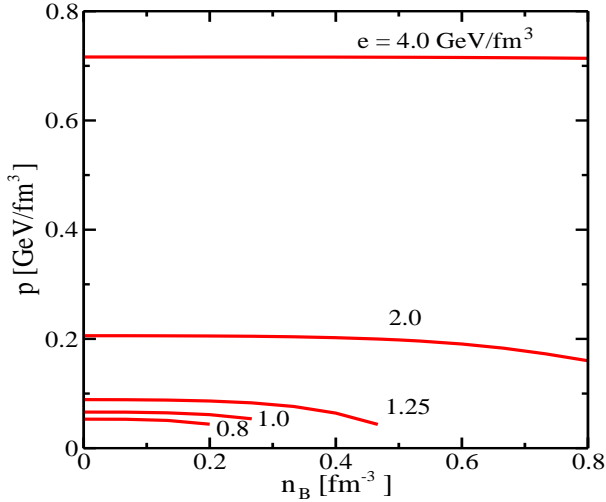}
  \caption{(Color online)
    Baryon number density dependence of the EoS $p(e,n_B)$ at constant 
    energy density $e$ as indicated. 
%The lines correspond (from top to bottom) to
%Left panel: $e=$ 2.0, 1.25, 1.0, 0.8 GeV fm$^{-3}$ (from top to bottom). 
    The curves end where the solution 
    of the flow equation (\ref{e:flow}) is not longer unique. 
%    Right panel: $e=$ 8.0, 4.0, 2.0, 1.25 GeV fm$^{-3}$ 
%    (from top to bottom). For larger energy densities 
%    the horizontal section extends more and more. 
    \label{fig:Ndep}}
\end{figure}
%%%%%%%%%%%%%%%%%%%%%%%%%%%%%%%%%%%%%%%%%%%%%%%%%%%%%%%%%%%%%%%%%%%%%%%%%%%%%%%
%
dependence of the pressure at fixed energy density. Figure~\ref{fig:Ndep}
shows that significant baryon density dependence of the pressure at fixed 
energy density arises only for $e\le 2$ GeV/fm$^3$. 
%(top panel). 
At the smallest energy densities considered here, the dependence of $p$ 
on $n_B$ cannot be determined over the entire $n_B$ region shown since
the flow equation (\ref{e:flow}) for $G^2(T,\mu_B)$ has no unique 
solution at large $\mu_B$ for temperatures far below the estimated 
transition temperature $T_c(\mu_B)$ \cite{Bormio}. However, in the 
family of equations of state that we will construct and employ in the 
following, this peculiar feature for small $e$ will not occur. Larger 
baryon densities which become relevant at AGS and CERN/SPS energies or 
the future CBM project at the FAIR/SIS300 facility deserve separate 
studies. Under RHIC and LHC conditions finite baryon density effects 
on the equation of state can be safely neglected at all energy densities 
for which the QPM model can be applied.

%%%%%%%%%%%%%%%%%%%%%%%%%%%%%%%%%%%%%%%%%%%%%%%%%%%%%%%%%%%%%%%%%%%%%%%%%%%%%
\subsection{Robustness of the QPM EoS $\bm{p(e,n_B\approx0)}$}
\label{sec3c}
%%%%%%%%%%%%%%%%%%%%%%%%%%%%%%%%%%%%%%%%%%%%%%%%%%%%%%%%%%%%%%%%%%%%%%%%%%%%%

\begin{figure}[ht]
  \includegraphics[bb=80 0 580 710,width=5cm,height=7.6cm,angle=-90.,clip=]%
                  {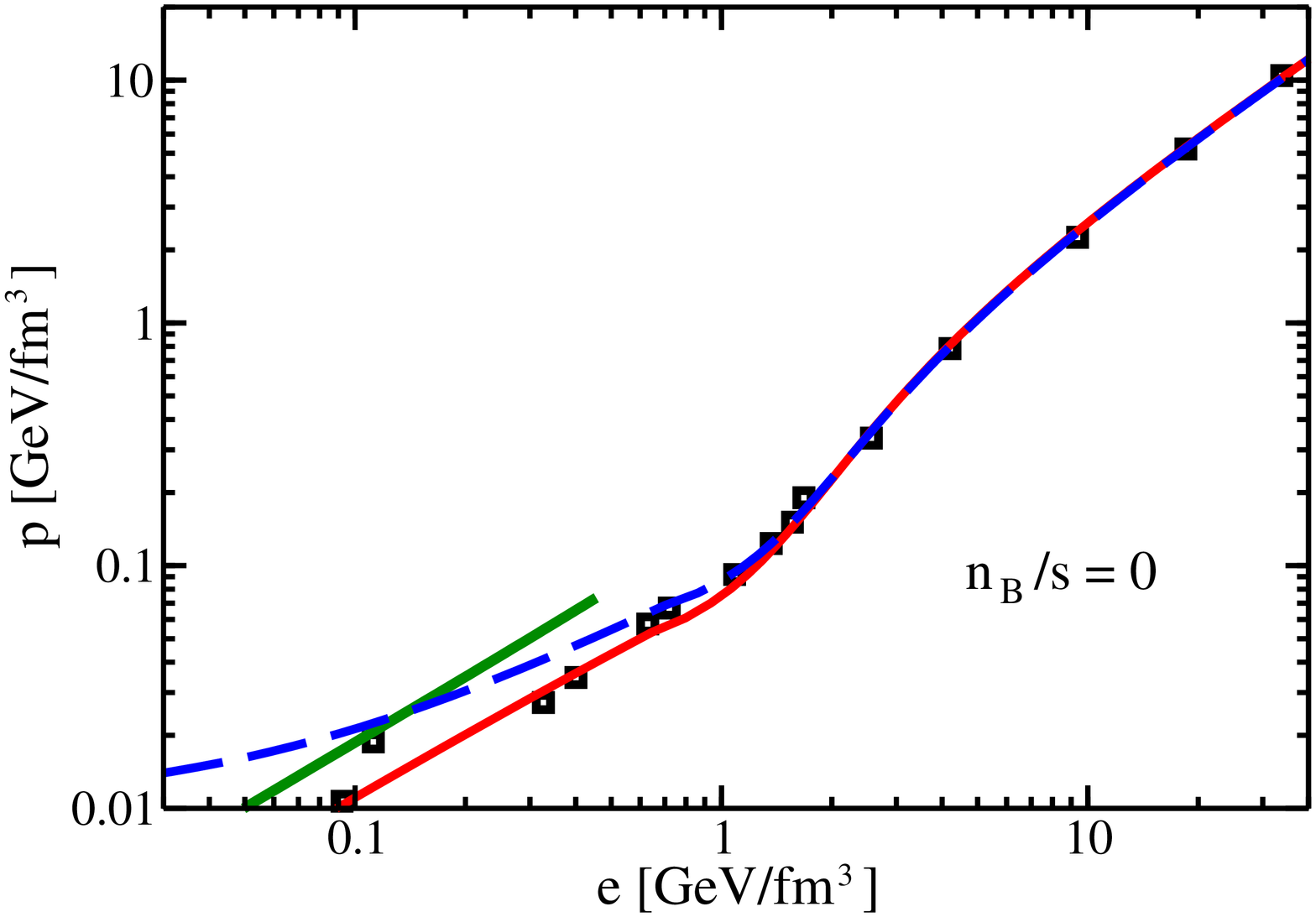}
  \includegraphics[bb=80 5 580 710,width=5cm,height=7.5cm,angle=-90.,clip=]%
                  {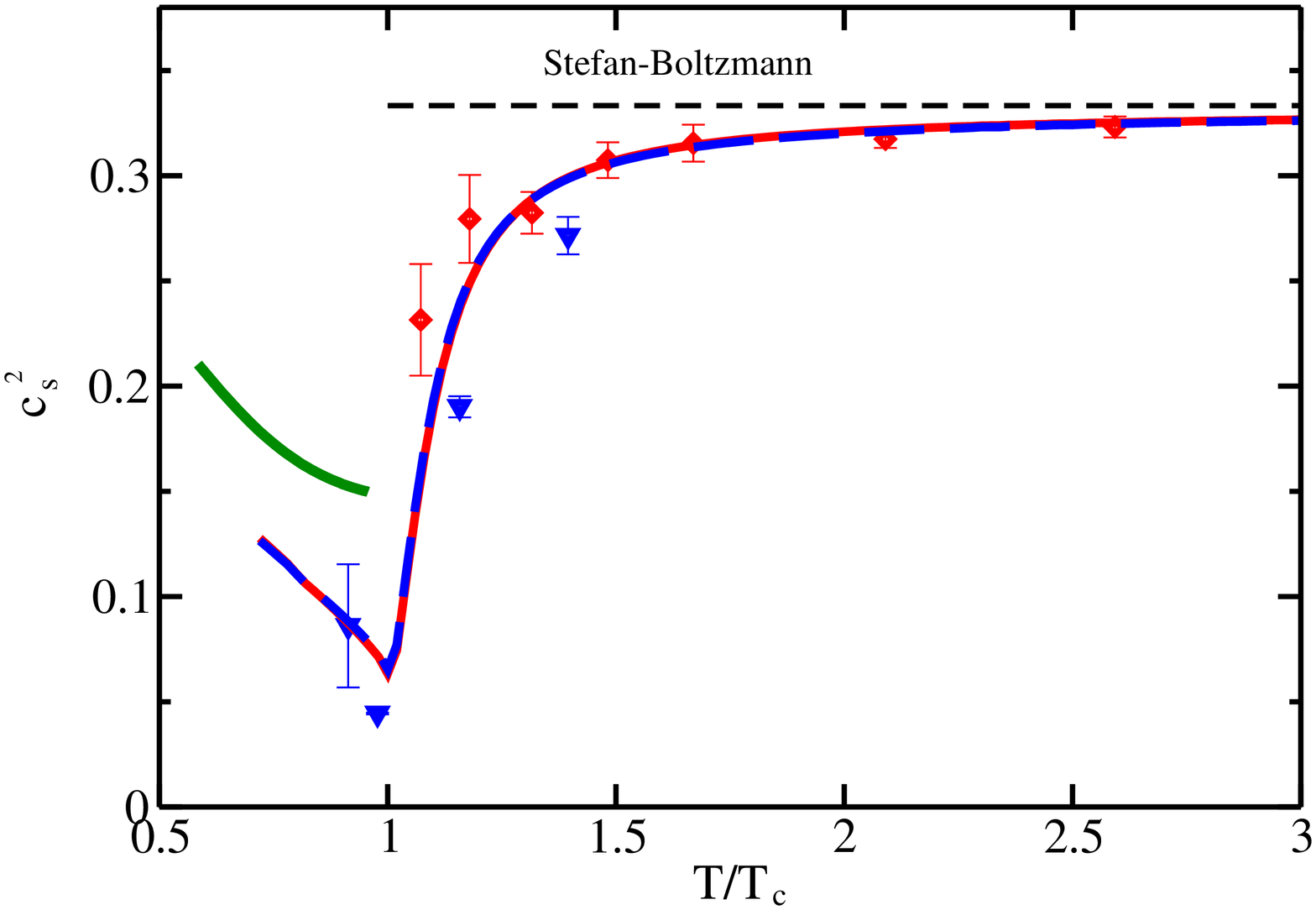}
  \includegraphics[bb=80 5 580 710,width=5cm,height=7.5cm,angle=-90.,clip=]%
                  {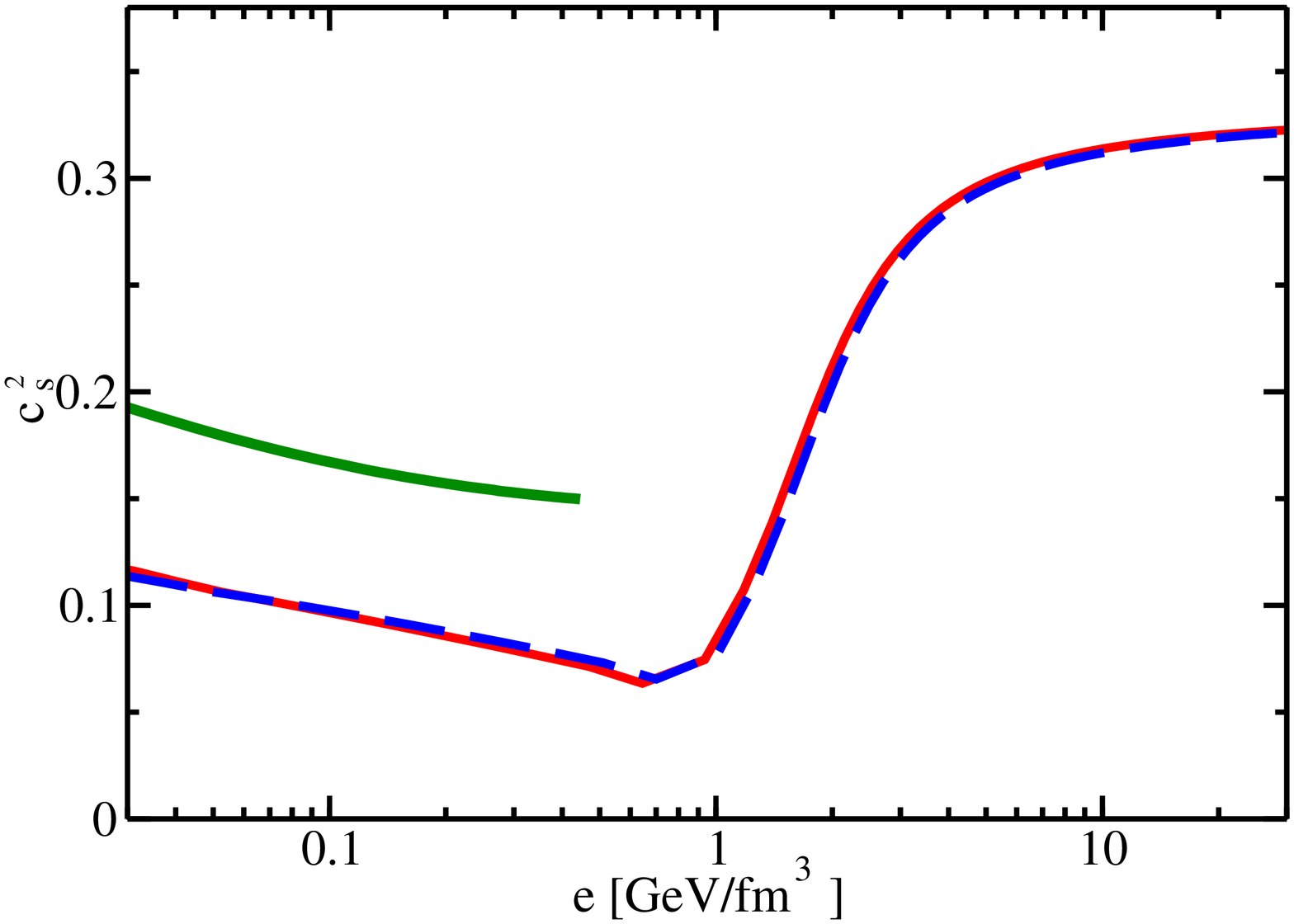}
  \caption{(Color online)
    {\sl Top panel:} QPM EoS for $N_f=2{+}1$ (solid red) and its 
    chiral extrapolation to physical quark masses (dashed blue). Squares 
    show LQCD data for $N_f=2+1$ quark flavors with unphysical masses
    \cite{Kar2}.  
    {\sl Middle panel:} Comparison of the squared speed of sound 
    $c_s^2 = \partial p/\partial e$ as a function of $T/T_c$
    from the QPM with lattice QCD data \cite{Kar4} (diamonds and triangles) 
    deduced from the $N_f=2$ data for $p(e)$ in \cite{Ejiri}. Differences 
    between the QPM fit to the LQCD data (solid red) and its 
    extra\-po\-lation to physical quark masses (dashed blue) for 
    $N_f=2{+}1$ are almost invisible. 
    {\sl Bottom panel:} Same as middle panel, but plotted as a function of 
    energy density $e$. -- In all three panels the solid green 
    line shows the hadron resonance gas model EoS ``aa1'' from \cite{AZHYDRO}. 
    \label{fig:EoS2}}
\end{figure}

We now perform a naive chiral extrapolation of the QPM EoS by setting 
$m_{q0}=0$ and $m_{s0}=150$\,MeV in the thermodynamic expressions, leaving 
all other parameters fixed. The resulting EoS is shown in the top panel of
Fig.~\ref{fig:EoS2}.
%
%%%%%%%%%%%%%%%%%%%%%%% Fig. 11 %%%%%%%%%%%%%%%%%%%%%%%%%%%%%%%%%%%%%%%%%%%%%%%
%%%%%%%%%%%%%%%%%%%%%%%%%%%%%%%%%%%%%%%%%%%%%%%%%%%%%%%%%%%%%%%%%%%%%%%%%%%%%%%
%
In this procedure a possible dependence of the QPM parameters in 
Eqs.~(\ref{e:G2param}), (\ref{e:G2pert}) and, especially, of the integration 
constant $B(T_c)$ in Eq.~(\ref{e:pres0}) on the quark mass parameters $m_{a0}$ 
is completely neglected. Note that in the transition region 
($e\sim 1$\,GeV/fm$^{3}$) the chirally extrapolated result exceeds the 
original QPM equation of state (which was fitted to lattice data with 
unphysical quark masses) by approximately 10\%. For higher energy 
densities $e\ge 2$ GeV/fm$^3$ these quark mass effects are seen to be 
negligible.

For $e\le 0.45$ GeV/fm$^{3}$, the fat solid line in the top panel of 
Fig.~\ref{fig:EoS2} shows a hadron resonance gas model EoS with a 
physical mass spectrum in chemical equilibrium \cite{AZHYDRO}. Obviously, 
it exceeds both the lattice QCD data and their QPM parametrization.
The chirally extrapolated QPM EoS, on the other hand, approaches 
and interesects the hadron resonance gas EoS. 

Considering $p/e$ as a function of $e$, we find for the lattice-fitted 
QPM EoS a softest point $(p/e)_\mathrm{min}=0.075$ at 
$e_c=0.92$\,GeV/fm$^{3}$. For the chirally extrapolated QPM EoS, the 
softest point moves slightly upward to $(p/e)_\mathrm{min}=0.087$ at 
$e_c=1.1$\,GeV/fm$^{3}$, in good agreement with the lattice QCD data 
which show a softest point $(p/e)_\mathrm{min}=0.080$ at 
$e_c=1$\,GeV/fm$^{3}$. 

The small differences between the lattice-fitted QPM equation of state and 
its chirally extrapolated version for $N_f=2{+}1$ can be further analyzed 
by studying the squared speed of sound $c_s^2$. In the middle panel of 
Fig.~\ref{fig:EoS2}, $c_s^2$ is shown as a function of $T/T_c$ for both 
versions of the QPM EoS and compared with lattice QCD results \cite{Kar4}. 
One sees that, as far as $c_s^2$ is concerned, the extrapolation of the 
QPM to physical quark masses has no discernible consequences, and both 
versions of the QPM EOS therefore have identical driving power for 
collective hydrodynamic flow. Hydrodynamically it is thus of no 
consequence that the available lattice QCD data for the EoS were obtained 
with unphysical quark masses.

\begin{figure}[ht]
  \includegraphics[bb=80 0 580 710,scale=0.34,angle=-90.,clip=]{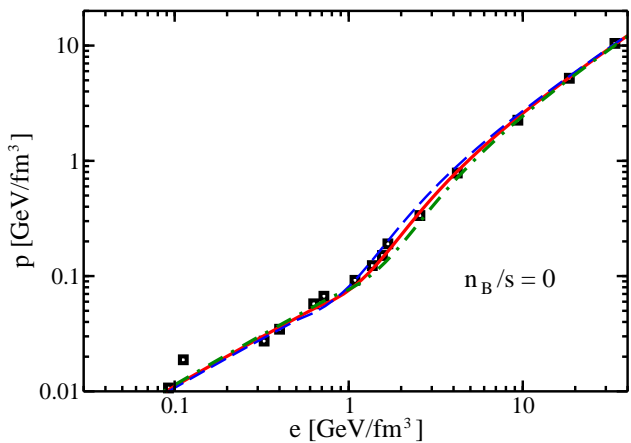}
  \caption{(Color online) 
    Dependence of the EoS for $N_f=2{+}1$ on the chosen value 
    of the physical scale $T_c$. Dashed, full and dash-dotted curves 
    correspond to $T_c=160$, $170$ and $180$ MeV, respectively. 
    Lattice data (squares) from \cite{Kar2}. 
  \label{fig:Tcdep}}
\end{figure}

The found EoS is also fairly robust against variations in the particular 
choice of the physical scale $T_c$. In Fig.~\ref{fig:Tcdep} we show $p(e)$ 
when setting $T_c=160$, $170$, and $180$ MeV, respectively, thereby covering
the ``reasonable range'' advocated in \cite{Peik,Karsch01}. For small energy 
densities and, in particular, for large $e\ge 5$ GeV/fm$^3$ the EoS is 
rather independent of the choice of the value for $T_c$. At intermediate 
$e$, $p(e)$ varies at most by $\pm20\%$ for $\Delta T_c=\pm10$ MeV. 
As discussed below (Section~\ref{sec:Interpol}), we must anyhow bridge 
over this intermediate region when interpolating between the QPM and
hadron resonance EoS, so this weak dependence on the physical scale 
$T_c$ is irrelevant in practice.

%
%%%%%%%%%%%%%%%%%%%%%%% Fig. 12 %%%%%%%%%%%%%%%%%%%%%%%%%%%%%%%%%%%%%%%%%%%%%%%
\begin{figure}[h]
  \includegraphics[bb=80 0 580 710,scale=0.34,angle=-90.,clip=]{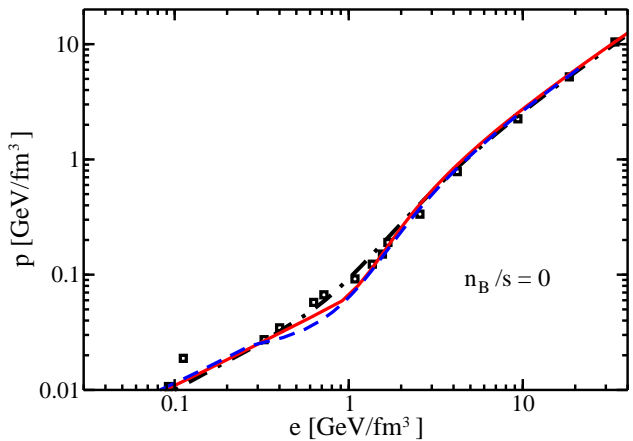}
  \caption{(Color online)
    Dependence of the EoS for $N_f=2$ on the employed continuum extrapolation
    as performed in Fig.~\ref{fig:c0para}. Dash-dotted, dashed and solid 
    curves correspond to the QPM parameterizations of the raw lattice 
    QCD data \cite{Kar1} and continuum extrapolations of these data by 
    a factor $d=1.1$ and $d=1.25$, respectively. 
  \label{fig:EoSc0para}}
\end{figure}

%%%%%%%%%%%%%%%%%%%%%%%%%%%%%%%%%%%%%%%%%%%%%%%%%%%%%%%%%%%%%%%%%%%%%%%%%%%%%%%
%
Next we examine variations in $p(e,n_B{\approx}0)$ arising from different 
continuum extrapolations of the lattice QCD data. Considering the 
various ``by hand'' continuum extrapolations of $p(T)/T^4$ shown in 
Fig.~\ref{fig:c0para} for $N_f=2$, the resulting EoS are plotted in 
Fig.~\ref{fig:EoSc0para}. Again, some weak sensitivity is observed only
in the transition region which will be bridged over in the next subsection
by matching the QPM EoS to a realistic hadron resonance gas below $T_c$. 
%
%%%%%%%%%%%%%%%%%%%%%%% Fig. 13 %%%%%%%%%%%%%%%%%%%%%%%%%%%%%%%%%%%%%%%%%%%%%%%
%%%%%%%%%%%%%%%%%%%%%%%%%%%%%%%%%%%%%%%%%%%%%%%%%%%%%%%%%%%%%%%%%%%%%%%%%%%%%%%
%
The problem discussed in section~\ref{sec:QPMcompNf2}, that different optimum 
QPM parameters are found by fitting the model to $c_0(T)$ or $c_2(T)$ 
(see Figs.~\ref{fig:2},~\ref{fig:truncs} and~\ref{fig:c0para}), does not 
matter here since the differences in the resulting parametrizations manifest
themselves only weakly in the EoS $p(e)$ and are completely negligible for 
$e>5$ GeV/fm$^{3}$. In the transition region near $e\approx 1$ GeV/fm$^3$ 
the resulting uncertainties are of order 20\% (see Fig.~\ref{fig:EoSc0para}), 
but again the interpolation to the hadronic EoS largely eliminates this
remaining sensitivity.

%
%%%%%%%%%%%%%%%%%%%%%%% Fig. 14 %%%%%%%%%%%%%%%%%%%%%%%%%%%%%%%%%%%%%%%%%%%%%%%
\begin{figure}[ht]
  \includegraphics[bb=50 15 580 710,scale=0.338,angle=-90.,clip=]{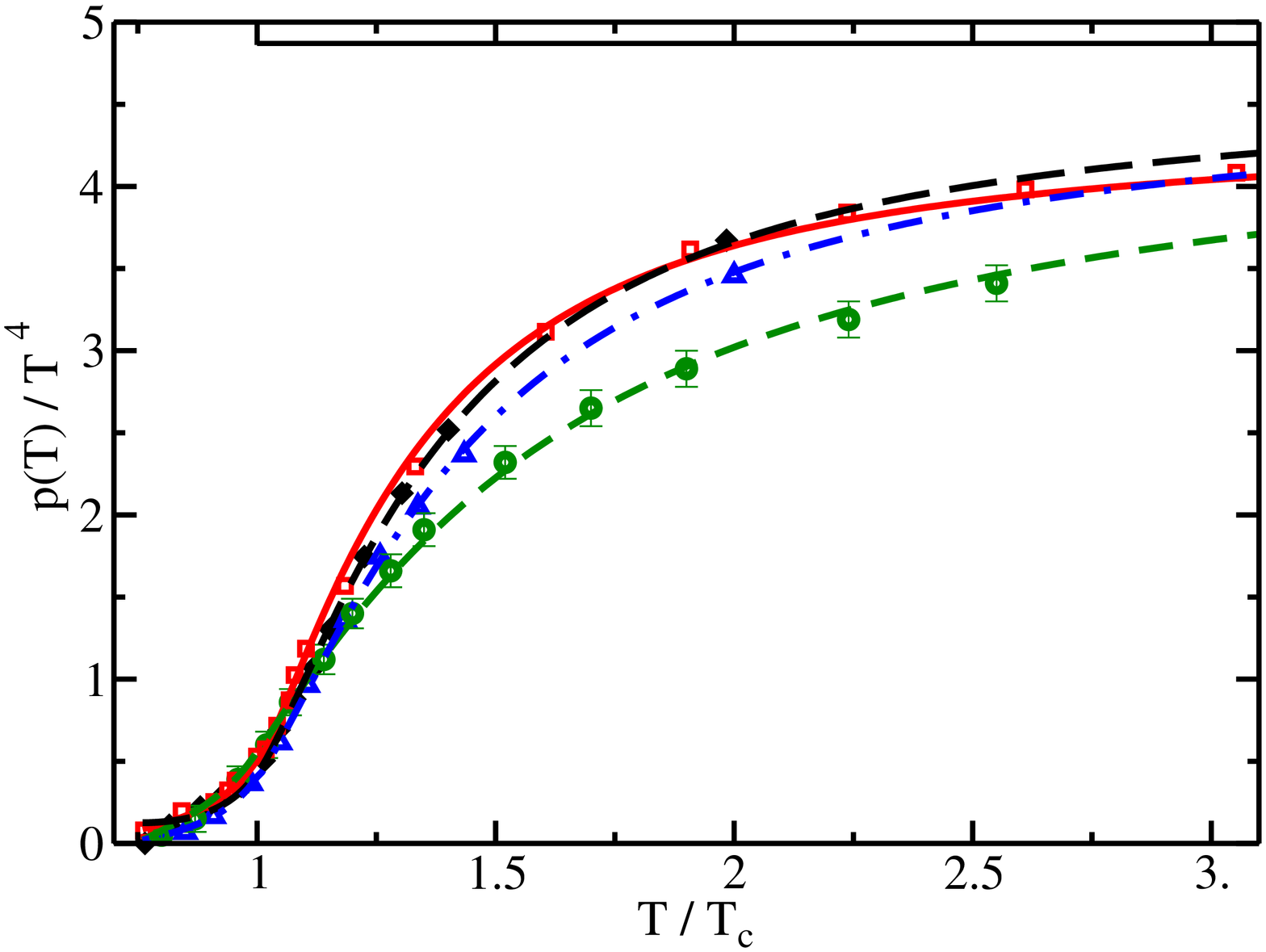}
  \includegraphics[bb=80 0 580 710,scale=0.345,angle=-90.,clip=]{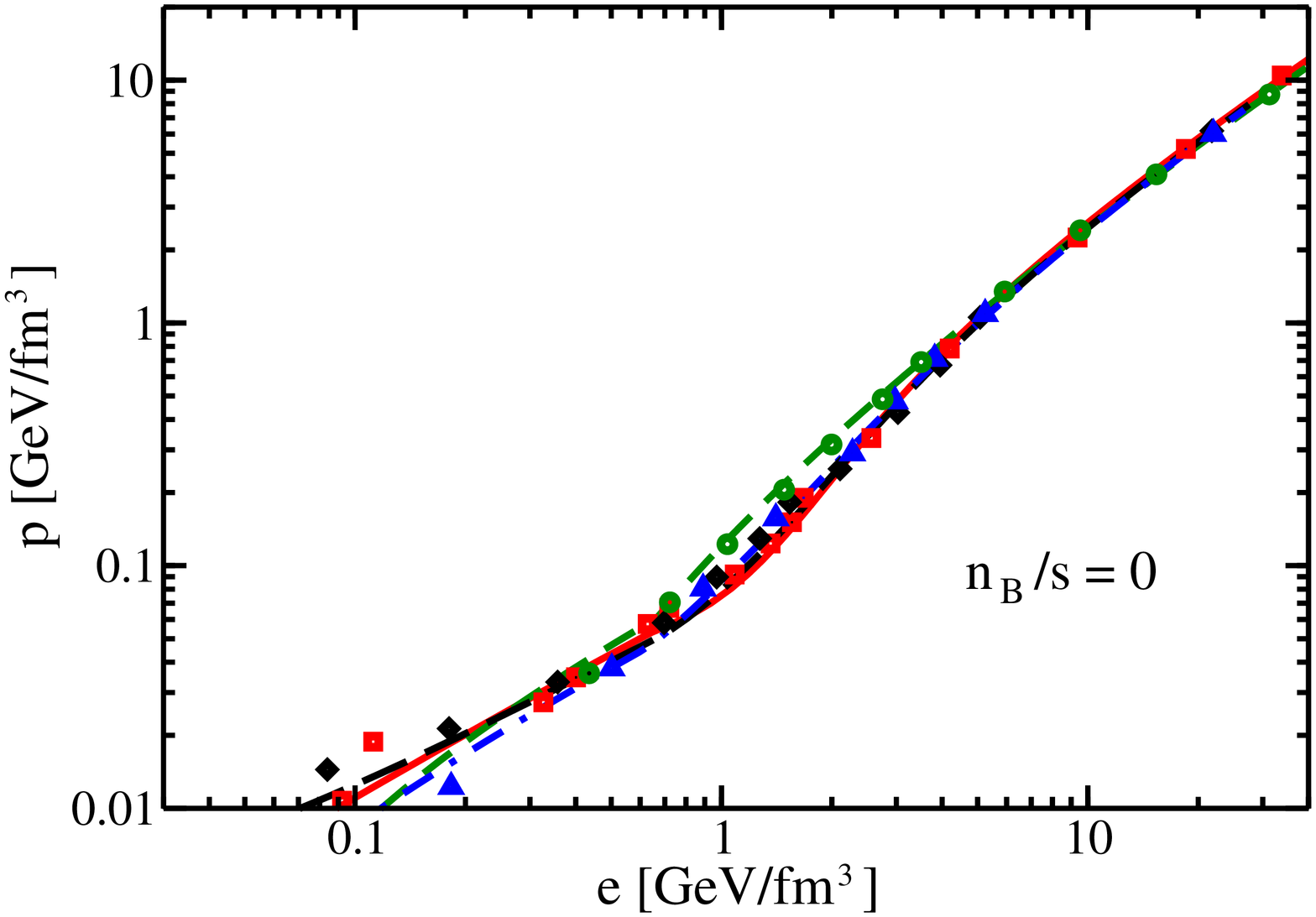}
  \caption{(Color online)
    Stability of the QPM EoS fitted to lattice QCD results for 
    $N_f{\,=\,}2{+}1$. 
    {\sl Top panel:} The scaled pressure $p(T)/T^4$ at $\mu_B{=}0$ from 
    different lattice QCD calculations (Ref.~\cite{Kar2} (squares), 
    Ref.~\cite{Bernard} (diamonds and triangles), and Ref.~\cite{Fodorlat3}
    (circles)), together with corresponding QPM fits (solid, long-dashed and 
    dash-dotted, and short-dashed lines, respectively). The fit parameters are 
    optimized separately in each case, keeping, however, $B(T_c)=0.51 T_c^4$ 
    with $T_c=170$\,MeV in all three parametrizations fixed. 
    {\sl Bottom panel:} The EoS $p(e,n_B{=}0)$ corresponding to the 
    data and fits shown in the top panel.
  \label{fig:difflat}}
\end{figure}
%%%%%%%%%%%%%%%%%%%%%%%%%%%%%%%%%%%%%%%%%%%%%%%%%%%%%%%%%%%%%%%%%%%%%%%%%%%%%%%
%

We close this subsection by exploring the robustness of the EoS $p(e)$
against variations between different existing lattice QCD simulations 
resulting from present technical limitations. In doing so we keep in 
mind the negligibly small baryon density effects in the region 
$n_B < 0.5$ fm$^{-3}$ pointed out above. In the top panel of 
Fig.~\ref{fig:difflat} we show the available lattice QCD results 
for $p(T)/T^4$ with $N_f{\,=\,}2{+}1$ dynamical quark flavors from
three different groups \cite{Kar2,Bernard,Fodorlat3} and compare them
with our QPM adjusted individually to each of these data sets. The 
differences between the data sets reflect the use of different lattice
actions, lattice spacings, bare quark masses etc. As shown in the figure,
these differences can be absorbed by the QPM through slight variations
in the fit parameters. However, when presenting the lattice results 
in the form of an EoS $p(e)$, they all coincide for $e\ge 5$\,GeV/fm$^{3}$ 
(bottom panel of Fig.~\ref{fig:difflat}). [The agreement is excellent
up to $e\approx 30$\,GeV/fm$^{3}$ while at higher energy densities a 
small difference of about 6\% between the equations of state from 
\cite{Kar2} and \cite{Fodorlat3} begins to become visible.] In this 
region the EoS can be parameterized by $p = \alpha e + \beta$ with 
$\alpha = 0.310 \pm 0.005$ and $\beta = -(0.56 \pm 0.07)$ GeV/fm$^3$. 
This robustness of the lattice QCD EoS for $e\ge 5$\,GeV/fm$^{3}$ implies 
that it can be considered as stable input for hydrodynamic simulations of 
heavy-ion collisions, and that the equation of state is well constrained 
at high energy densities. Our effort to substitute the often used bag 
model EoS above $T_c$ by a realistic QPM EoS which incorporates the 
lattice data seems therefore well justified.

%%%%%%%%%%%%%%%%%%%%%%%%%%%%%%%%%%%%%%%%%%%%%%%%%%%%%%%%%%%%%%%%%%%%%%%%%%%%%%
\subsection{Matching lattice QCD to a hadron resonance gas equation
of state via the QPM} 
\label{sec:Interpol}
%%%%%%%%%%%%%%%%%%%%%%%%%%%%%%%%%%%%%%%%%%%%%%%%%%%%%%%%%%%%%%%%%%%%%%%%%%%%%%

In this subsection we will now match the lattice QCD EoS at high energy
densities with a realistic hadron resonance gas model at low energy 
densities \cite{Hagedorn,Solf}. Since available lattice QCD simulations 
still employ unrealistic quark masses while the hadron gas model builds 
upon the measured spectrum of hadronic resonances, we will use the QPM 
to parametrize the lattice QCD EoS and extrapolate it to physical quark 
masses. Such quark mass effects matter most at the lower end of the 
temperature range covered by the lattice QCD data which is, however, also 
the region where the transition from the QPM to the hadron resonance gas 
model must be implemented.

In the vicinity of the phase transition, the conditions of the lattice 
QCD evaluations in Refs.~\cite{Kar1,Ejiri} correspond to a pion mass 
$m_\pi\approx 770$ MeV. This large pion mass reduces the pressure 
at small energy density below that of a realistic hadron resonance 
gas. Smaller quark masses are necessary to properly account for the
partial pressure generated by the light pion modes and their remnants 
in the temperature region around $T_c$. On the other hand, the hadron 
resonance gas model has been shown to be consistent with the QCD lattice 
data below $T_c$ if one appropriately modifies its mass spectrum for 
consistency with the employed lattice parameters \cite{Kar2}. We will 
therefore adopt the hadron resonance gas model with physical mass 
spectrum \cite{Hagedorn,Solf} as an appropriate approximation of the 
hadronic phase \cite{Berni}, and use the QPM to parametrize the lattice 
QCD EoS near and above $T_c$.

For the hadron resonance gas EoS \cite{Hagedorn,Solf} we use the 
implementation developed for the (2+1)-dimensional hydrodynamic code 
package AZHYDRO \cite{AZHYDRO} which provides this EoS in tabulated form
on a grid in the $(e,n_B)$ plane. Specifically, we use EoS ``aa1'' from 
the OSCAR website \cite{AZHYDRO} up to $e_1 = 0.45$ GeV/fm$^{3}$. It
describes a thermalized, but chemically non-equilibrated hadron resonance 
gas, with hadron abundance yield ratios fixed at all temperatures 
at their chemical equilibrium values at $T=T_c=170$\,MeV, as found 
empirically \cite{PBM} in Au+Au collisions at RHIC.

As seen in Fig.~\ref{fig:EoS2}, the pressure $p(e)$ of the hadron 
resonance gas EoS does not join smoothly to that of the QPM EoS at 
$T_c$ (i.e. at $e_1=0.45$\,GeV/fm$^3$), irrespective of whether one uses
directly the QPM fit to the lattice QCD data with unphysical quark masses
(solid red line in Fig.~\ref{fig:EoS2}) or extrapolates the QPM to 
physical quark masses (dashed blue line). A thermodynamically consistent 
treatment thus requires a Maxwell like construction, equating the two pressures 
at a common temperature $T_c$ and baryon chemical potential $\mu_B$. We opt 
here for a slightly different approach which
%is not fully consistent thermodynamically but 
has the advantage of allowing a systematic 
exploration of the effects of details 
(e.g., stiffness or velocity of sound)
of the EoS near $T_c$ on 
hydrodynamic flow patterns: We interpolate $p(e,n_B)$ at fixed baryon 
density $n_B$ linearly between the hadron resonance gas (``aa1'') value 
at $e=e_1$ to its value in the QPM at a larger value $e_\mathrm{m}$, 
keeping $e_1$ fixed but letting the ``matching point'' value $e_\mathrm{m}$ 
vary. In our procedure $T(e_\mathrm{m})\geq T(e_1)$, so $T(e)$ is also
interpolated linearly, as is the baryon chemical potential $\mu_B(e)$
at fixed $n_B$. (This is a convenient pragmatic procedure to interpolate
the special tabular forms of the EoS between $e_1$ and $e_m$ employed below.
Complete thermodynamic consistency would require involved polynomials
for temperature and chemical potential interpolation.
We utilize the linearized structures since the hydrodynamical evolution
equations do not explicitly refer to $T$ and $\mu_B$ in the interpolation
region; instead, only $p(e,n_B)$ matters.)  

\begin{figure}[ht]
  \includegraphics[bb=20 20 363 280,scale=0.75,angle=0.,clip=]%
                  {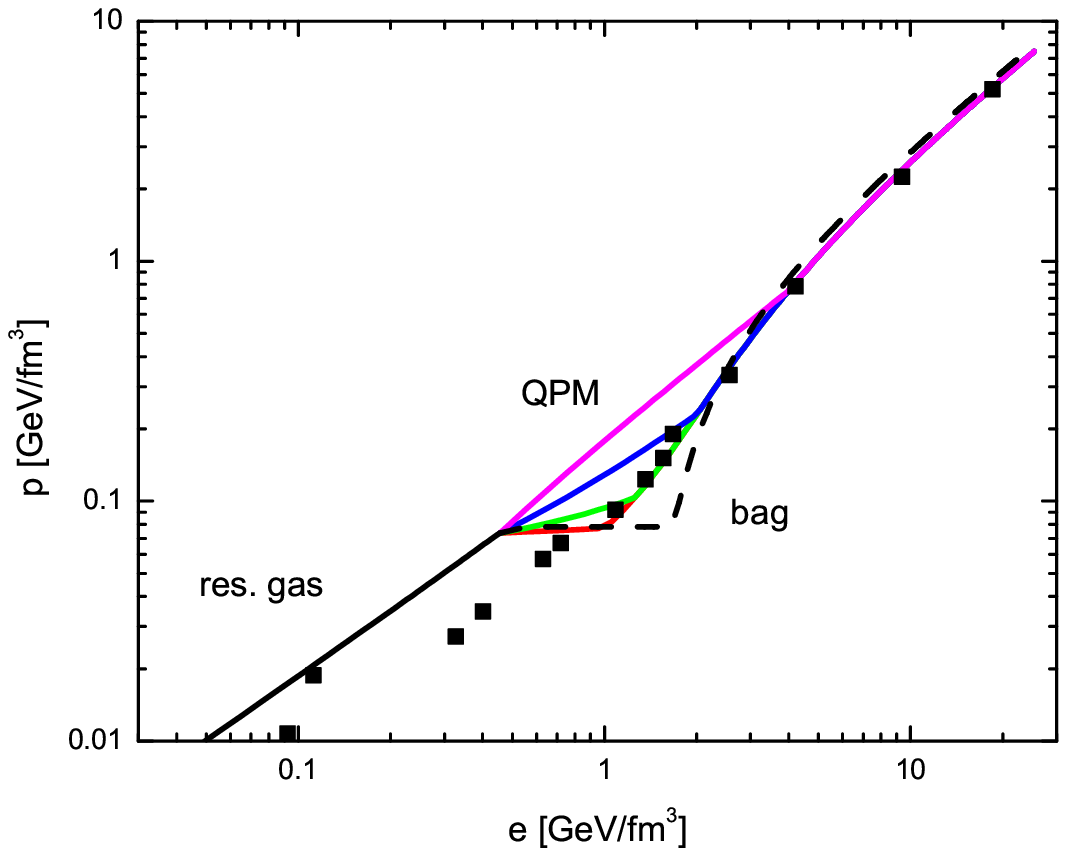}
  \includegraphics[bb=16 20 363 280,scale=0.75,angle=0.,clip=]%
                  {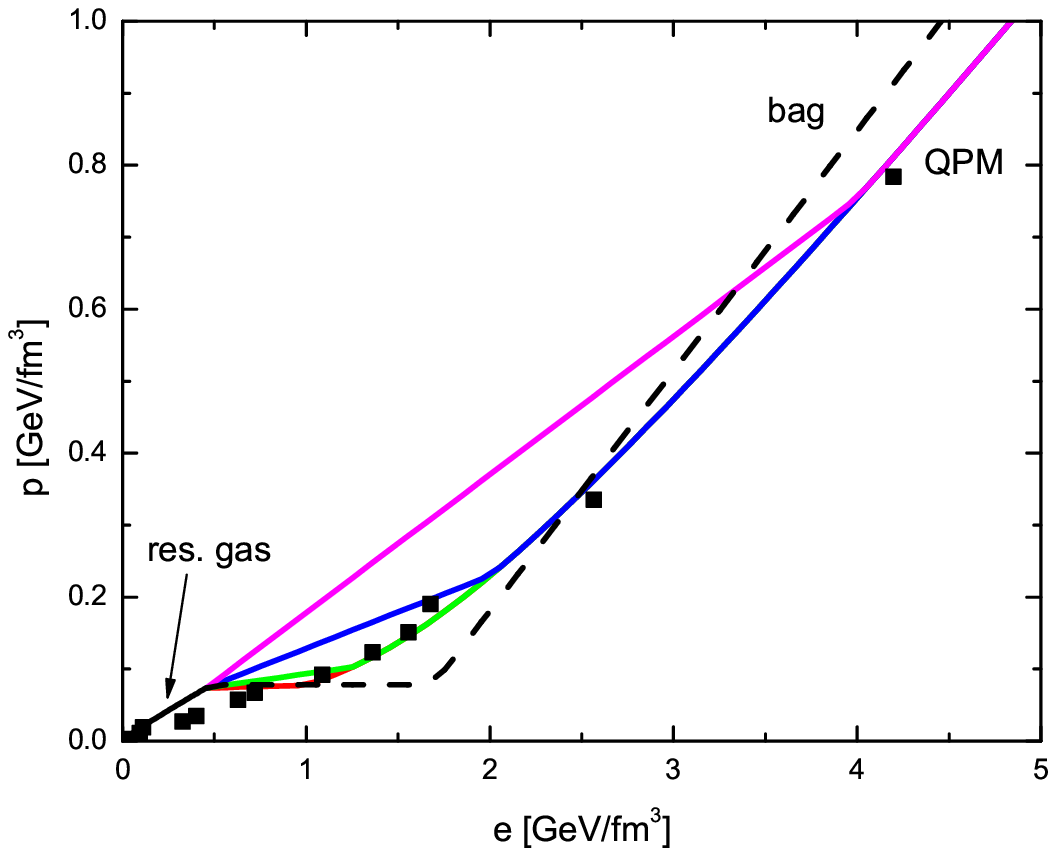}
  \caption{(Color online)
    A family of equations of state for $N_f=2+1$, combining our 
    QPM at high energy densities with a hadron resonance gas model 
    (``res. gas'') in the low energy density regime through linear 
    interpolation. We show the range of energy densities relevant 
    for collisions at RHIC. The solid lines show $p(e)$ for QPM(4.0), 
    QPM(2.0), QPM(1.25), and QPM(1.0) (from top to bottom), where the
    numerical label indicates the matching point $e_\mathrm{m}$ in 
    GeV/fm$^3$. On the given scale, effects of varying $n_B$ between 0 
    and 0.5\,fm$^{-3}$ are not visible. Lattice QCD data (squares) are
    from Ref.~\protect{\cite{Kar2}}. For comparison a bag model EoS (``bag'') 
    with a sharp first order phase transition is also shown (dashed line). The bottom
    panel zooms in onto the transition region, using a linear energy 
    density scale.
    \label{fig:Family}}
\end{figure}

This produces a family of equations of state whose members are
labelled by the matching point energy density $e_\mathrm{m}$. We here
explore the range $1.0$\,GeV/fm$^3 \leq e_\mathrm{m} \leq 4.0$\,GeV/fm$^3$
(see Fig.~\ref{fig:Family}). Since the chiral extrapolation of the QPM fit
to physical quark masses significantly affects the EoS $p(e)$ only at 
energy densities below 1\,GeV/fm$^3$ (see top panel in Fig.~\ref{fig:EoS2}), 
it does not matter whether we use for this procedure the direct QPM fit to 
the lattice QCD data or its chiral extrapolation. 

%
%%%%%%%%%%%%%%%%%%%%%%%%%% Fig. 15 %%%%%%%%%%%%%%%%%%%%%%%%%%%%%%%%%%%%%%%%%%%
%%%%%%%%%%%%%%%%%%%%%%%%%%%%%%%%%%%%%%%%%%%%%%%%%%%%%%%%%%%%%%%%%%%%%%%%%%%%%%%
%

Figure~\ref{fig:Family} shows the result for four selected $e_\mathrm{m}$
values, $e_\mathrm{m}=1.0,\, 1.25,\, 2.0,$ and 4.0\,GeV/fm$^3$
(from bottom to top). For $e_\mathrm{m}{\,=\,}3.0$\,GeV/fm$^3$ one obtains a 
curve $p(e)$ (not shown) that extrapolates the hadron resonance gas with 
constant slope all the the way to the QPM curve. The dashed line in 
Fig.~\ref{fig:Family} shows a Maxwell construction between the hadron 
resonance gas and a bag model equation of state with $c_s^2=\frac{1}{3}$; 
this results in a strong first order phase transition with latent heat 
$\Delta e_\mathrm{lat} = 1.1$\,GeV/fm$^3$ (``EoS\,Q'' in~\cite{KSH,AZHYDRO}). 

Our construction differs from the approach explored in \cite{TLS}
where the hadron resonance gas is matched to an ideal quark-gluon gas
with varying values for the latent heat $\Delta e_{lat}$. For example, 
varying the latent heat in EoS~Q from $\Delta e_{lat}=0.4$\,GeV/fm$^3$ 
to 0.8 and $1.6$\,GeV/fm$^3$, the pressure $p(e_0,n_B{\,=\,}0)$ at a 
typical initial energy density $e_0{\,=\,}30$\,GeV/fm$^3$ for central 
Au+Au collisions at RHIC decreases by only 1.4\% and 4.3\%, respectively, 
with correspondingly small changes in the entropy density $s_0$. In 
our approach, however, 
the entropy density $s_0$ at $e_0$ is given by lattice QCD and 
significantly ($\sim 15\%$) smaller. We note that our QPM(1.0) is
similar to EOS\,Q in \cite{KSH,AZHYDRO}, except for the larger latent
heat of EoS~Q.

%
%%%%%%%%%%%%%%%%%%%%%%%%%% Fig. 16 %%%%%%%%%%%%%%%%%%%%%%%%%%%%%%%%%%%%%%%%%%%%
\begin{figure}[ht]
  \includegraphics[bb=20 20 363 280,scale=0.78,angle=0.,clip=]{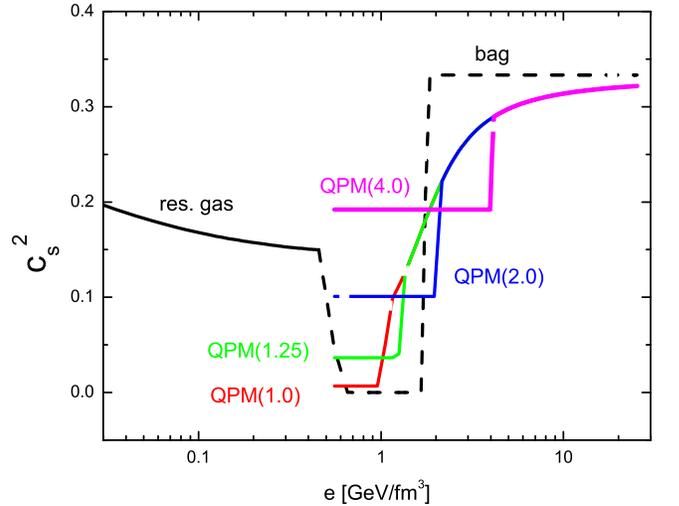}
  \caption{(Color online) 
    Squared speed of sound $c_s^2 = \partial p/\partial e$ as a function 
    of energy density $e$ along an isentropic expansion trajectory with 
    $s/n_B=100$, for the EoS family QPM($e_m$) depicted in 
    Fig.~\ref{fig:Family}. Baryon density effects are not visible on the 
    given scale as long as $n_B < 0.5$\,fm$^{-3}$.
  \label{fig:sound}}
\end{figure}
%%%%%%%%%%%%%%%%%%%%%%%%%%%%%%%%%%%%%%%%%%%%%%%%%%%%%%%%%%%%%%%%%%%%%%%%%%%%%%%
%

Figure \ref{fig:sound} shows the corresponding squared speed of sound,
$c_s^2$, as a function of energy density $e$. The linear interpolation
between the hadron resonance gas at $e\leq e_1=0.45$\,GeV/fm$^3$ and the 
QPM at $e\geq e_\mathrm{m}$ leads to a region of constant sound speed for
$e_1 \leq e \leq e_\mathrm{m}$. This constant increases monotonically
with the matching point value $e_\mathrm{m}$. For $e_\mathrm{m}=3$\,GeV/fm$^3$,
the hadron resonance gas extrapolates smoothly to the QPM, with no ``soft
region'' of small sound speed left over at all. In this case the typical
phase transition signature of a softening of the EoS near $T_c$ is minimized,
leading to minimal phase transition effects on the development of 
hydrodynamic flow.
 
%%%%%%%%%%%%%%%%%%%%%%%%%%%%%%%%%%%%%%%%%%%%%%%%%%%%%%%%%%%%%%%%%%%%%%%%%%%%%%%
\section{Azimuthal Anisotropy and Transverse Momentum Spectra}
\label{sec:ellipticflow}
%%%%%%%%%%%%%%%%%%%%%%%%%%%%%%%%%%%%%%%%%%%%%%%%%%%%%%%%%%%%%%%%%%%%%%%%%%%%%%%

Equipped with our QCD based family of equations of state, we can now
explore the effects of fine structures in the EoS near $T_c$ on the
evolution of hydrodynamic flow, by computing the transverse momentum
spectra $dN/(dy\,p_T dp_T\,d\phi)$ and elliptic flow $v_2(p_T)$ for a variety
of hadron species. To emphasize flow effects, we only consider directly 
emitted hadrons and neglect resonance decay distortions.

In non-central heavy-ion collisions, the initial almond shaped cross 
section of the overlap zone perpendicular to the beam direction in 
coordinate space is converted into an azimuthally asymmetric momentum 
distribution due to the appearance of a radially non-symmetric flow 
governed by pressure gradients. Assuming no transverse flow at a certain 
``initial time'' $\tau_0$, at which the hydrodynamical expansion stage 
starts, the azimuthal asymmetry is determined by the acting pressure. 
Therefore, the azimuthal asymmetry is an ideal probe of the equation of 
state. In addition, the final anisotropy in the momentum distribution 
depends on the rescatterings among the particles and serves as measure 
of the degree of local thermalization. 

The asymmetry is quantified by the harmonic coefficients of an expansion 
of the emitted hadrons transverse momentum spectra into a Fourier series 
in the azimuthal emission angle $\phi$ around the beam axis relative to 
the reaction plane (which is determined by the direction of the impact 
parameter $b$): 
\begin{equation}
  \label{e:spectrum}
  \frac{dN}{p_\perp dp_\perp dy \,d\phi} = 
  \frac{dN}{2\pi \,p_\perp dp_\perp dy}\left(1 + 
  2\,v_2(p_\perp ,y)\cos{2\phi} + \dots \,\right) .
\end{equation}
The second Fourier coefficient $v_2(p_\perp ,y) = 
\langle\cos{2\phi}\rangle_{p_\perp ,y}$ is called elliptic flow. We here
exploit the 2+1 dimensional relativistic hydrodynamic program package with 
Cooper-Frye freeze-out formalism, AZHYDRO, used in Refs.~\cite{KSH,%
Huovinen,Huov05,Hirano01}. It assumes longitudinally boost-invariant 
expansion \`a la Bjorken. Clearly, this is appropriate only near
midrapidity $y\approx 0$, but sufficient for purposes of our qualitative
investigation here.

Different phenomenological equations of state of strongly interacting 
matter were proposed in previous studies \cite{Solf,KSH,TLS,Shu,Huovinen,%
Huov05,Hirano01,Teaney_chem,Rapp_chem,KR03}, exhibiting either a strong
first order phase transition with different values of latent heats 
\cite{Solf,KSH,TLS,Huovinen,Hirano01,KR03}, a smooth but rapid crossover 
\cite{Huov05}, or no phase transition at all \cite{Solf}. 
These equations of state differ significantly in their high-density 
regions and softest points, and in the speed of sound which controls 
details of the developing flow pattern. Investigating the hydrodynamic
consequences of different equations of state helps to establish benchmarks 
for tracing specific phase transition signatures and distinguishing them 
from other dynamical features (such as so far poorly explored viscous 
effects). 

We emphasize, however, that we do not attempt here a systematic comparison 
with RHIC data. Previous studies \cite{KSH,TLS,Huov05} have already 
qualitatively established that existing data are best described by an EoS 
with a phase transition or rapid crossover of significant strength (i.e.
featuring a strong increase of the entropy and energy density within a 
narrow temperature interval) that exhibits both a soft part near $T_c$ and
a hard part not too far above $T_c$. More quantitative statements about a 
preference of one form of the EoS over another require a discussion 
that goes beyond the pure ideal fluid dynamical approach discussed here, 
due to well-known strong viscous effects on the evolution of elliptic flow 
in the late hadron resonance gas phase \cite{Hirano:2005xf}. Studying the 
effects of EoS variations within a more complete framework that allows to 
account for non-ideal fluid behaviour in the very early and late stages of 
the fireball expansion is an important task for the future. Staying here 
within the ideal fluid approach, we do note, however, that our discussion 
improves over that presented in \cite{Huov05} by employing below $T_c$ a 
{\em chemically non-equilibrated} hadron resonance gas EoS which correctly 
reproduces the measured hadron yields, irrespective of the selected value 
for the hydrodynamic decoupling temperature. 

%%%%%%%%%%%%%%%%%%%%%%%%%%%%%%%%%%%%%%%%%%%%%%%%%%%%%%%%%%%%%%%%%%%%%%%%%%%%
\subsection{Top RHIC Energy}
%%%%%%%%%%%%%%%%%%%%%%%%%%%%%%%%%%%%%%%%%%%%%%%%%%%%%%%%%%%%%%%%%%%%%%%%%%%%

We employ P.~Kolb's program package version 0.0 available from the OSCAR 
archive \cite{AZHYDRO}. While the study presented in \cite{KSH} shows that 
at RHIC energies ($\sqrt{s}\sim 200\,A$\,GeV) most of the finally observed
momentum anisotropy develops before the completion of the quark-hadron 
phase transition, the build-up of elliptic flow still occurs mostly in 
the temperature region where the lattice QCD data show significant 
deviations from an ideal quark-gluon gas. It is therefore of interest to
investigate the effects of these deviations, and of variations of the exact
shape of the EoS in the transition region, on the final elliptic flow in 
some detail, both at RHIC energies, where they are expected to matter, and
at higher LHC energies where most (although not all \cite{Hirano:2007xd})
of the anisotropic flow will develop before the system enters the phase 
transition region, thus reducing its sensitivity to the transition region.

We fix the initial conditions for top RHIC energy according to \cite{KSH}
\begin{equation}
s_0 = 110\, {\rm fm}^{-3}, \quad
n_0 = 0.4\, {\rm fm}^{-3}, \quad
\tau_0 = 0.6\, {\rm fm/c};
\label{RHIC_init}
\end{equation} 
these parameters describing the initial conditions in the fireball 
center for central ($b{\,=\,}0$) Au+Au collisions are required input 
for the hydro code \cite{AZHYDRO}. From these initial conditions
for central collisions the initial profiles for non-central collisions 
are calculated using the Glauber model \cite{KSH}. For our EoS these 
values translate (independently of the QPM version used) into 
$e_0 = 29.8$\,GeV/fm$^3$, $p_0 = 9.4$\,GeV/fm$^3$, and $T_0 = 357$\,MeV. 
[Strictly speaking, since in the QPM the physical scale is set by $T_c$, 
varying $T_c$ in the range $170 \pm 10$ MeV would result in a variation 
of $e_0$ from 25 to 33 GeV/fm$^3$ when keeping $s_0$ fixed (such as to 
maintain the same final charged particle multiplicity 
$dN_{ch}/dy \propto s_0 \tau_0$). We fix $T_c=170$\,MeV.] 

%
%%%%%%%%%%%%%%%%%%%%%%%%%%%%%%%%%%%%% Fig. 17 %%%%%%%%%%%%%%%%%%%%%%%%%%%%%%%%%
\begin{figure}[ht]
  \includegraphics[bb= 20 20 350 280,scale=0.73,angle=0,clip=]%
                  {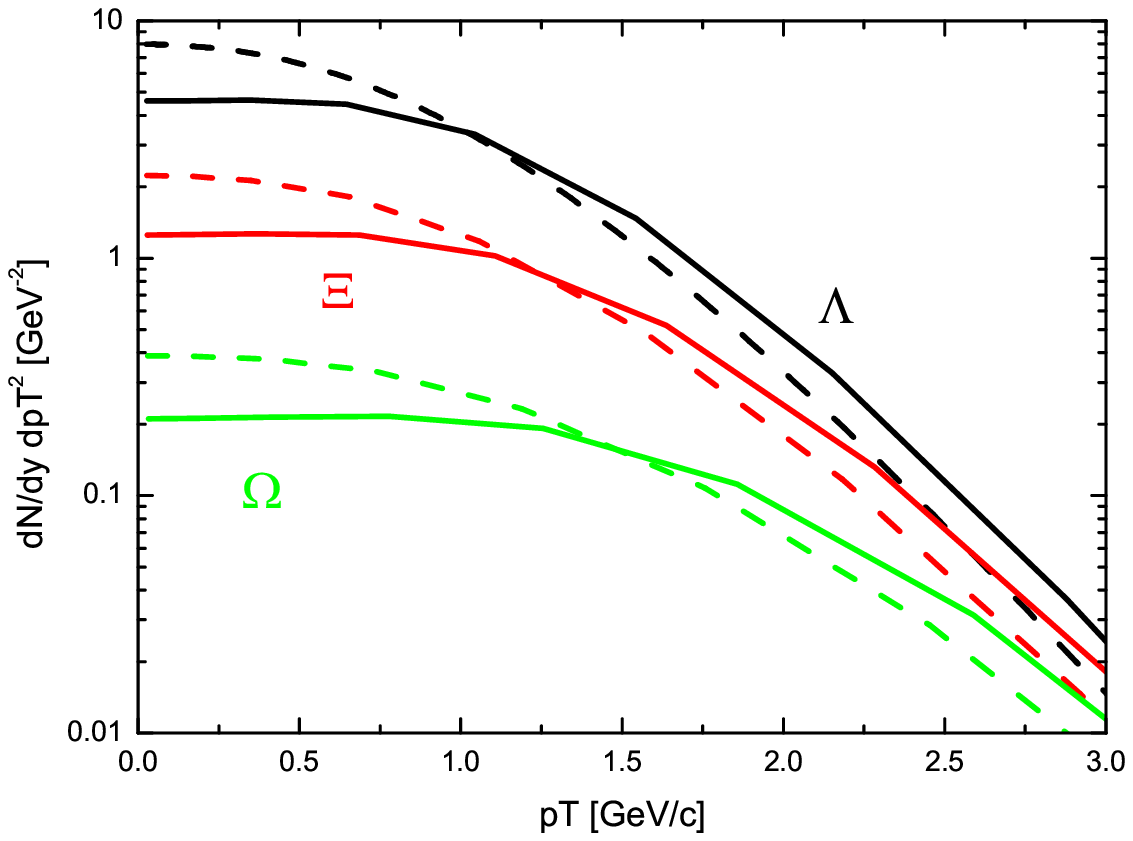}
  \includegraphics[bb= 20 20 330 280,scale=0.73,angle=0,clip=]%
                  {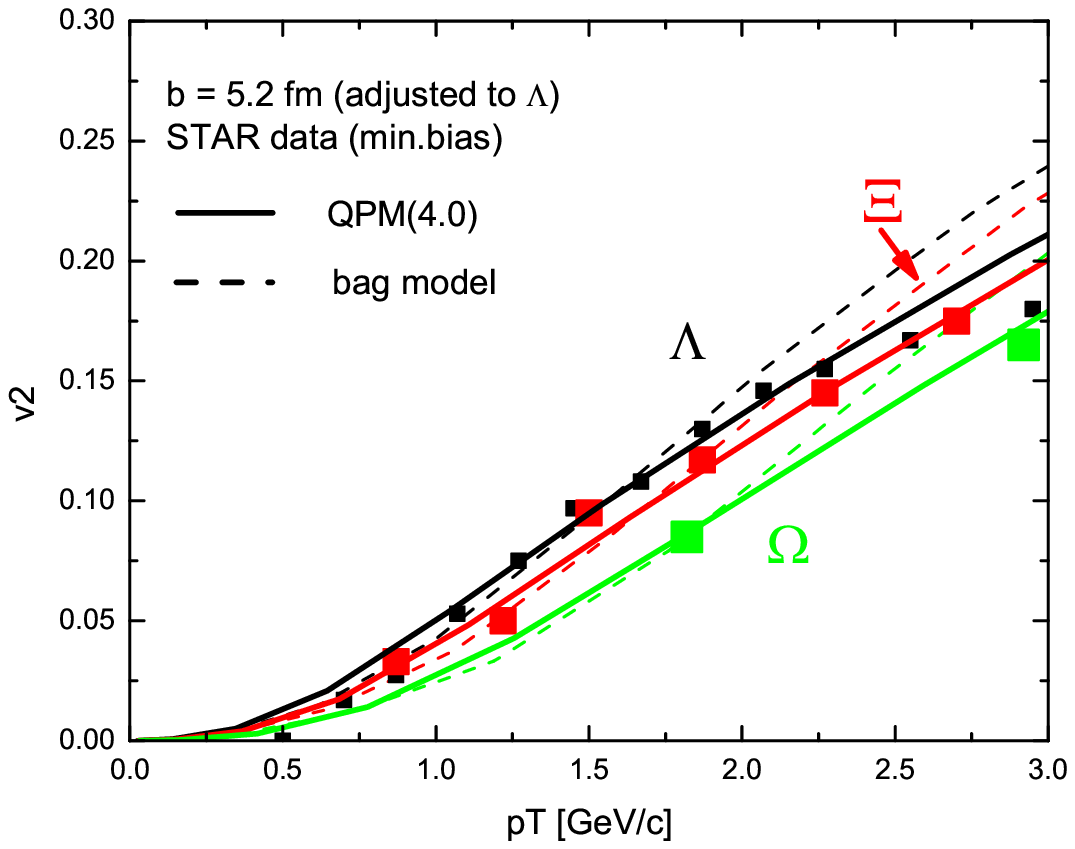}  
  \caption{(Color online)
  Transverse momentum spectrum (left) and elliptic flow coefficient
  (right) for directly emitted strange baryons. The symbols 
  represent STAR data \cite{STAR_multistrange} (see text for details). 
  Solid and dashed curves are for EoS QPM(4.0) and the bag model EoS, 
  respectively.
  \label{RHIC}}
\end{figure}
%%%%%%%%%%%%%%%%%%%%%%%%%%%%%%%%%%%%%%%%%%%%%%%%%%%%%%%%%%%%%%%%%%%%%%%%%%%%%%%
%

Our calculations assume zero initial transverse velocity, $v_{\perp 0} = 0$
at $\tau=\tau_0$. In the hadron phase, the Rapp-Kolb chemical 
off-equilibrium EoS \cite{KR03} is used to account for frozen-out chemical 
reactions. The freeze-out criterion is $e_\mathrm{f.o.} = 0.075$\,GeV/fm$^3$, 
corresponding to a freeze-out temperature of about 100 MeV. All hadrons
are assumed to freeze out at the same energy density.

The usual approach when analyzing data is to adjust the set of initial and 
final conditions to keep the transverse momentum spectra of a given set of 
hadron species fixed, and to then study the variation of $v_2$. Here we 
instead illustrate the impact of the EoS by using a fixed set of initial 
and freeze-out parameters. We explore Au+Au collisions at a fixed impact 
parameter $b=5.2$\,fm, adjusted to best reproduce minimum bias data from 
the STAR collaboration.

In Fig.~\ref{RHIC} we show the transverse momentum spectra and differential
elliptic flow for directly emitted $\Lambda$, $\Xi$, and $\Omega$ hyperons. 
These hadron species do not receive large resonance decay contributions, so 
by comparing the results for directly emitted particles with the measured
spectra one can obtain a reasonable feeling for the level of quality of
the model description. We show only results obtained with the two extreme 
equations of state, QPM(4.0) and the bag model EoS (see
Fig.~\ref{fig:Family}). The results for QPM(1.0) are very similar to those
from the bag model EoS, although the latter features a larger latent heat. 
The two remaining equations of state (QPM(1.25) and QPM(2.0)) interpolate   
smoothly between the extreme cases shown in Fig.~\ref{RHIC}.

The left panel in Fig.~\ref{RHIC} shows that QPM(4.0) generates significantly
larger radial flow, resulting in flatter $p_T$ spectra especially for the
heavy hadrons shown here. This can be understood from Fig.~\ref{fig:Family}
since this EoS does not feature a soft region with small speed of sound
around $T_c$. Flatter $p_T$ spectra generically result in smaller 
Fourier coefficients $v_2(p_T)$ \cite{KSH}, but the right panel in 
Fig.~\ref{RHIC} shows that for $p_T<1.5$\,GeV/$c$, QPM(4.0) actually 
produces larger $v_2(p_T)$ than the bag model EoS. This implies that 
QPM(4.0) also produces a larger overall momentum anisotropy 
(i.e. $p_T$-integrated elliptic flow) than the bag model EoS, again due 
to the absence of a soft region near $T_c$. Only at large $p_T>2$\,GeV/$c$, 
where the ideal fluid dynamic picture is known to begin to break down 
\cite{Heinz_SQM04}, does QPM(4.0) give smaller elliptic flow than the bag 
model EoS, as naively expected \cite{KSH} from the flatter slope of the 
single particle $p_T$-distribution.

The larger $v_2(p_T)$ at low $p_T<1.5$\,GeV/$c$ from QPM(4.0) is not 
favored by the data. In this sense we confirm the qualitative conclusion 
from earlier studies \cite{KSH,TLS,Huov05} that the data are best described
by an EoS with a soft region near $T_c$, followed by a rapid increase
of the speed of sound $c_s$ above $T_c$.

%  
%%%%%%%%%%%%%%%%%%%%%%%%%%% Fig. 18 %%%%%%%%%%%%%%%%%%%%%%%%%%%%%%%%%%%%%%%%%%%
\begin{figure*}[htb]
  \includegraphics[bb= 20 20 350 280,scale=0.73,angle=0.,clip=]%
                  {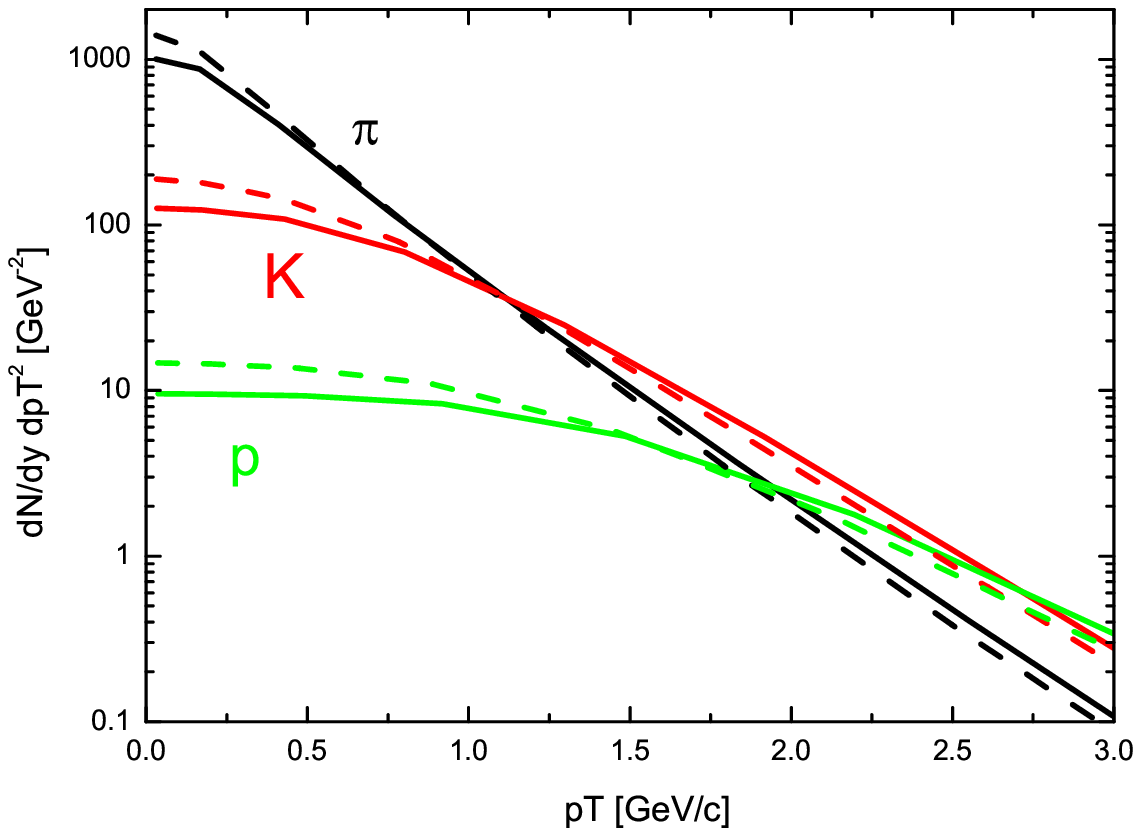}
  \includegraphics[bb= 20 20 350 280,scale=0.73,angle=0.,clip=]%
                  {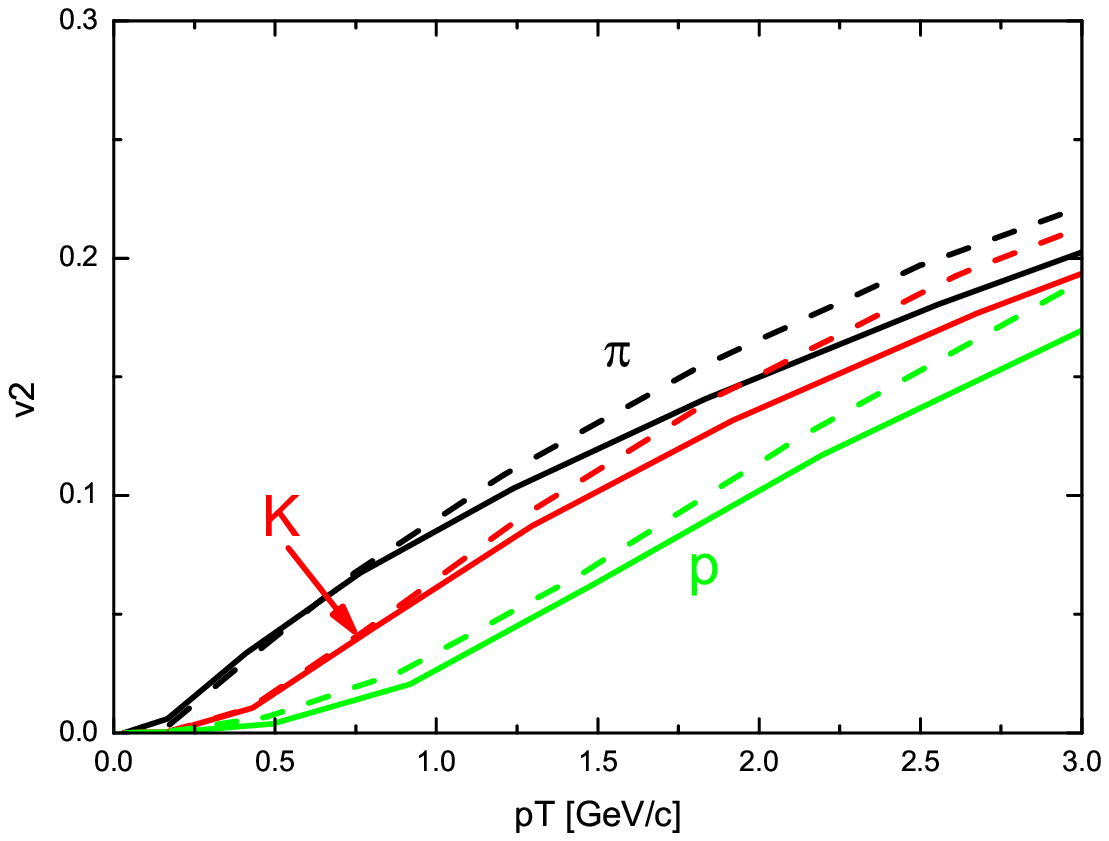}\\
  \includegraphics[bb= 20 20 350 280,scale=0.73,angle=0.,clip=]%
                  {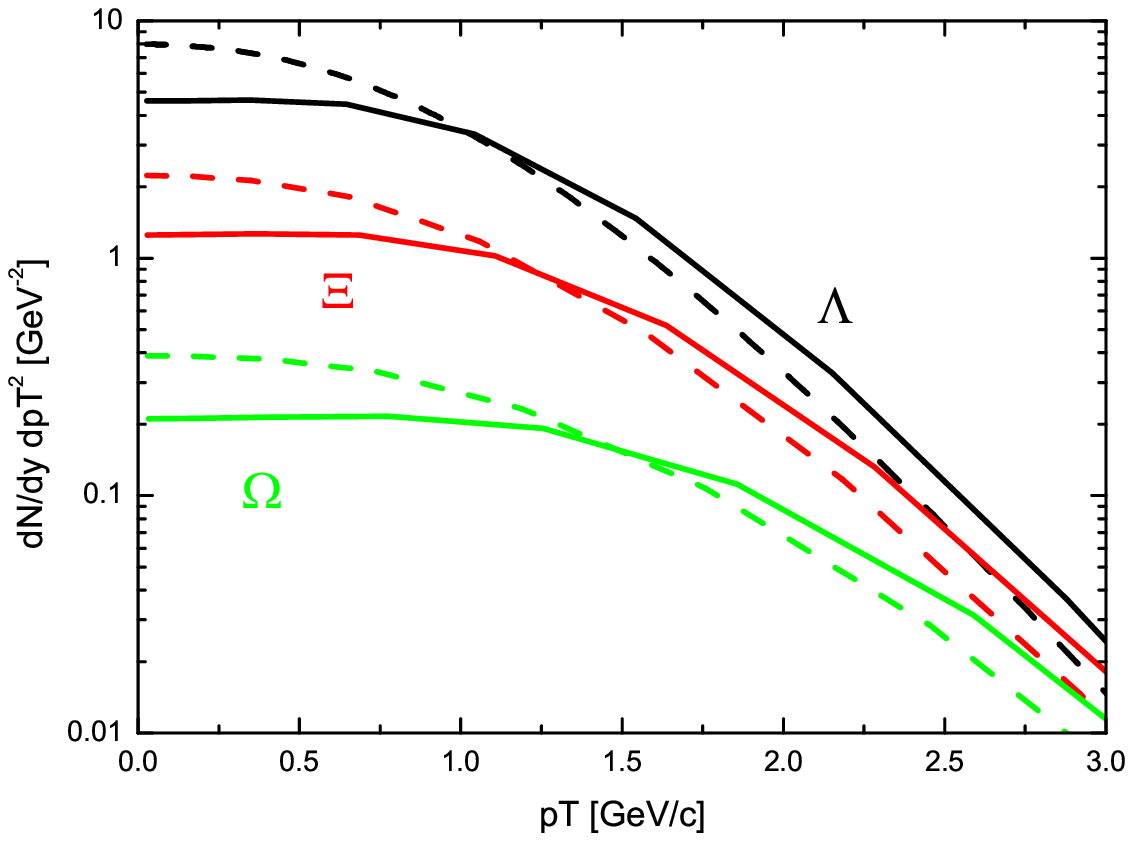}  
  \includegraphics[bb= 20 20 350 280,scale=0.73,angle=0.,clip=]%
                  {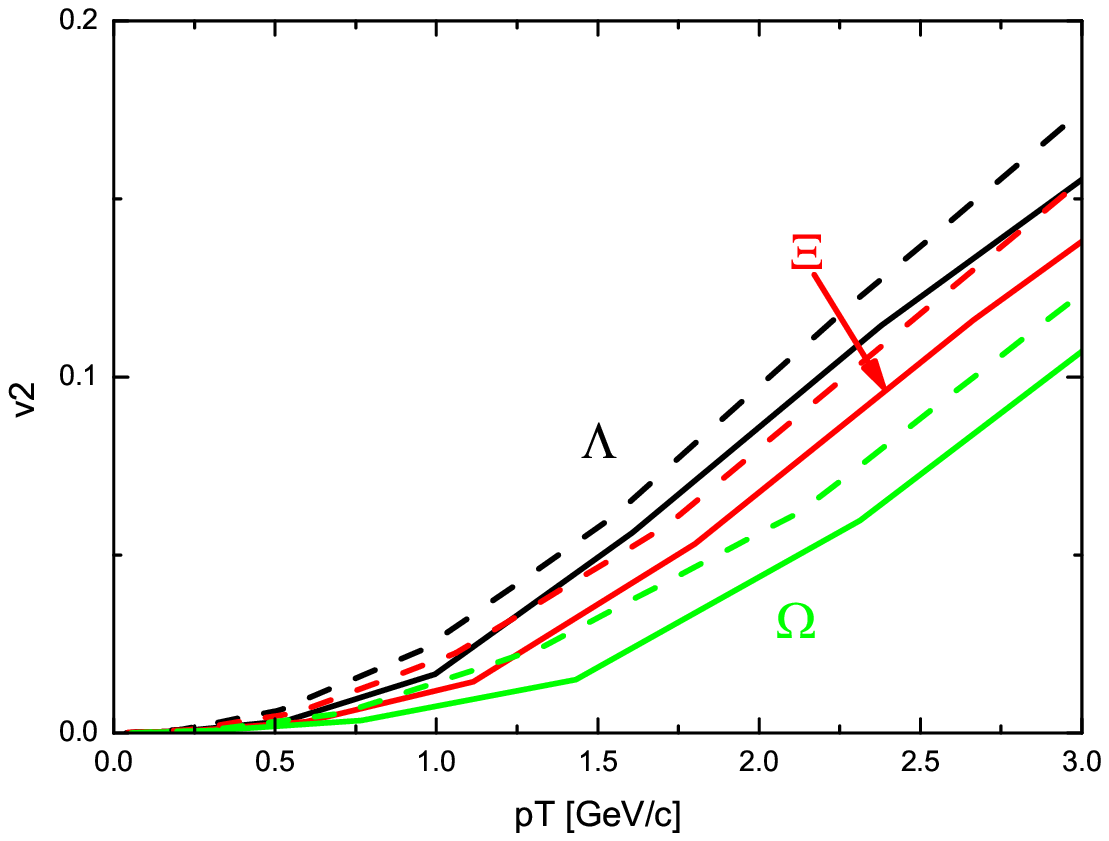} 
  \caption{(Color online)
  Transverse momentum spectrum (left panels) and azimuthal anisotropy 
  (right panels) for pions, kaons and protons (upper row) and strange 
  baryons (lower row). Initial conditions according to eq.~(\ref{LHC_init}). 
  The spectra show only directly emitted hadrons. Solid and dashed curves 
  are for EoS QPM(4.0) and the bag model EoS being similar to QPM(1.0), 
  respectively. 
  \label{LHC}}
\end{figure*}
%%%%%%%%%%%%%%%%%%%%%%%%%%%%%%%%%%%%%%%%%%%%%%%%%%%%%%%%%%%%%%%%%%%%%%%%%%%%%%%
%

%%%%%%%%%%%%%%%%%%%%%%%%%%%%%%%%%%%%%%%%%%%%%%%%%%%%%%%%%%%%%%%%%%%%%%%%%%%%%%%
\subsection{LHC estimates}
%%%%%%%%%%%%%%%%%%%%%%%%%%%%%%%%%%%%%%%%%%%%%%%%%%%%%%%%%%%%%%%%%%%%%%%%%%%%%%%

Predictions for Pb+Pb collisions at the LHC involve a certain amount of
guesswork about the initial conditions at the higher collision energy.
We here do not embark upon a systematic exploration of varying initial
conditions, as proposed e.g. in Refs.~\cite{LHC_init_conds}, but simply 
guess conservatively
\begin{equation}
s_0 = 330\,{\rm fm}^{-3}, \quad
\tau_0 = 0.6\,{\rm fm/c},
\label{LHC_init}
\end{equation} 
keeping all other parameters unchanged. This corresponds to 3 times larger
final multiplicities than measured at RHIC. Within the QPM these initial 
parameters translate into $e_0 = 127$\,GeV/fm$^3$, $p_0 = 42$\,GeV/fm$^3$, 
and $T_0 = 515$\,MeV for the peak values in central Pb+Pb collisions.
We again study collisions at impact parameter $b = 5.2$\,fm, using the
Glauber model to calculate the corresponding initial density profiles
from the above parameters.

Again we show results only for the two extreme equations of state, QPM(4.0)
and the bag model EoS. Generally, the $p_T$ spectra for LHC initial 
conditions are flatter than for RHIC initial conditions, since the higher
initial temperature and correspondingly longer fireball lifetime results
in stronger radial flow. Figure~\ref{LHC} shows that again QPM(4.0), which 
lacks a soft region near $T_c$, generates even larger radial flow (i.e. 
flatter $p_T$ spectra) than the bag model EoS (whose results are similar 
to those obtained with QPM(1.0)). The radial flow effects are particularly 
strong for the heavy hyperons. 

The overall momentum anisotropy (i.e. the $p_T$-integrated elliptic flow)
does not increase very much between RHIC and LHC \cite{KSH}. Since the
LHC spectra are flatter, i.e. have more weight at larger $p_T$ than 
the RHIC spectra, the elliptic flow {\em at fixed $p_T$} must therefore 
decrease. This is clearly seen when one compares the right panels of
Figs.~\ref{RHIC} and \ref{LHC}. The decrease is particularly strong for
the hyperons at low $p_T$ where the LHC transverse momentum spectra 
become extremely flat.

%%%%%%%%%%%%%%%%%%%%%%%%%%%%%%%%%%%%%%%%%%%%%%%%%%%%%%%%%%%%%%%%%%%%%%%%%%%%%%%
\section{Summary}
\label{sec:conclusions}
%%%%%%%%%%%%%%%%%%%%%%%%%%%%%%%%%%%%%%%%%%%%%%%%%%%%%%%%%%%%%%%%%%%%%%%%%%%%%%%

We have shown that available lattice QCD calculations give converging and 
robust results for the EoS $p(e,n_B)$ in the region of large energy density.
Baryon density effects were shown to be negligibly small for 
$n_B < 0.5$\,fm$^{-3}$, i.e. the EoS relevant for heavy ion collisions
at top RHIC and LHC energies is the same. In the transition region 
(i.e. for temperatures around $T_c$) different lattice calculations
still exhibit quantitative differences. The lattice calculations examined 
here do not yet join smoothly at low energy densities (i.e. at $T<T_c$) to 
the hadron resonance gas model EoS with physical mass spectrum. While our 
quasiparticle model covers all considered lattice QCD equations of state 
and serves as a reliable tool to connect thermodynamic quantities 
in a thermodynamically consistent way, it is not obvious that a reliable 
chiral extrapolation is feasible by simply replacing the quark mass 
parameters employed on the lattice by their physical values. If we do so
we find significant quark mass effects only for energy densities below
about 1\,GeV/fm$^3$, i.e. below the hadronization phase transition.
 
In the present paper we therefore assumed as a working hypothesis the 
validity of the hadron resonance gas model EoS below $T_c$ (i.e. below
an energy density of $e_1=0.45$\,GeV/fm$^3$) and interpolated this EoS 
linearly to the robust high energy density branch from the QPM fit to the
lattice QCD data. In doing so we arrive at a family of equations of state
whose members QPM($e_\mathrm{m}$) are labeled by the matching point 
energy density $e_\mathrm{m}$ where we join the QPM EoS. The resulting
equations of state QPM($e_\mathrm{m}$) are available in the usual 
tabulated form on the OSCAR website \cite{AZHYDRO}. We find that the 
uncertain intermediate region, which is bridged over by this 
interpolation procedure, has a small but non-negligible impact on the 
evolution of radial and elliptic flow in high energy heavy-ion collisions, 
visible in the transverse momentum spectra and elliptic flow coefficients
of various (directly emitted) hadron species. Existing RHIC data seem
to favor those members of our family of equations of state that exhibit
a soft region near $T_c$ followed by a rapid rise of the speed of sound
towards the ideal gas value above $T_c$. We caution, however, that we did 
not perform a systematic study including simultaneous variations of the 
EoS and initial and final conditions, and that event-by-event fluctuations 
\cite{Asa00a,Jeo00a,Koda1,Koda2} or viscous effects \cite{Teaney} may wash 
out differences between different sets of equations of state. More
quantitative conclusions about the EoS require systematic investigations 
which match the ideal fluid description to viscous dynamical models 
for the very early and late stages of the fireball expansion; this is
left for the future.    

\begin{acknowledgments}

This work was supported by BMBF 06DR121/06DR136, GSI-FE and Helmholtz VI, 
as well as by the U.S. Department of Energy under contract 
DE-FG02-01ER41190. We thank S. Fodor, S. Hands, P. Huovinen, F. Karsch, 
E. Laermann, A. Peshier, K. Redlich, and S. Wheaton for fruitful discussions.

\end{acknowledgments}

%%%%%%%%%%%%%% References %%%%%%%%%%%%%%%%%%%%%%%%%%%%%%%%%%%%%%%%%%%%%%%%%%%


\begin{thebibliography}{100}
\bibitem{fn1}
These parameters differ significantly from the QPM parameters that optimally 
reproduce $c_0(T)$ from Ref.~\cite{Kar1} (cf.\ Fig.~\ref{fig:2}). This is 
most likely a consequence of different algorithms and lattice QCD parameters
used in Refs. \cite{Kar1} and \cite{All05}. It implies that the dependence 
of these lattice data on the quark masses and lattice spacing, combined 
with the presently incompletely controlled \cite{Fodornew} continuum 
extrapolation, is too complex to be fully captured by our QPM. If we
compute the pressure $p(T,\mu_B{=}0)$ (i.e. $c_0(T)$) from the QPM
with the parameters used in Figs.~\ref{fig:truncs} and \ref{fig:coeffs2}
(which were tuned to $c_2(T)$ from Ref.~\cite{All05}), adjusting the extra 
parameter $B(T_c)$ which is not needed for the $c_n(T)$, $n\ge 2$, to the value
$B(T_c)=0.61 T_c^4$, we find a result that is about 10\% larger than that 
shown in Fig.~\ref{fig:2} for large $T$ and rises somewhat more rapidly
in the vicinity of $T_c$, resulting in a larger entropy density.
\end{thebibliography}
\end{document}